\documentclass[journal,a4paper]{IEEEtran}

\usepackage[T1]{fontenc}

\usepackage{graphicx}
\usepackage[caption=false,font=footnotesize]{subfig}

\usepackage{amsthm,amsmath,amssymb,bm}
\usepackage{mathrsfs}
\usepackage{amsfonts}

\newtheorem{theorem}{Theorem}

\newtheorem{lemma}{Lemma}

\newtheorem{remark}{Remark}

\newtheorem{proposition}{Proposition}

\usepackage{algorithm}
\usepackage{algpseudocode}

\usepackage{xcolor}
\usepackage{booktabs}
\usepackage{placeins}
\usepackage{verbatim}

\usepackage[colorlinks,linkcolor=blue,urlcolor=black,citecolor=blue]{hyperref}

\usepackage{siunitx}
\begin{document}
	
\title{\textcolor{black}{Model-Driven Deep Learning for Distributed Detection with Low-Bit Quantization}}
\author{Wei~Guo,~\IEEEmembership{Member,~IEEE},~Meng~He,~Chuan~Huang,~\IEEEmembership{Member,~IEEE},~Hengtao~He,~\IEEEmembership{Member,~IEEE},~Shenghui~Song,~\IEEEmembership{Senior Member,~IEEE},~Jun~Zhang,~\IEEEmembership{Fellow,~IEEE},~Khaled~B.~Letaief,~\IEEEmembership{Fellow,~IEEE}\vspace{-2.5em}
\thanks{	
	An earlier version of this paper was presented in part at the 2024 IEEE International Mediterranean Conference on Communications and Networking \cite{guo2024model}.

	W. Guo, S. Song, J. Zhang, and K. B. Letaief are with the Department of Electronic and Computer Engineering, \textcolor{black}{The Hong Kong University of Science and Technology}, Hong Kong SAR. Emails: \{eeweiguo, eeshsong, eejzhang, eekhaled\}@ust.hk.
	
	M. He is with the Future Network of Intelligence Institute, and School of Science and Engineering, The Chinese University of Hong Kong, Shenzhen 518172, China. Email: menghe@link.cuhk.edu.cn.

	C. Huang is with the Shenzhen Institute for Advanced Study, University of Electronic Science and Technology of China, and Future Network of Intelligence Institute, Shenzhen 518172, China. Email: huangch@uestc.edu.cn.
	
	H. He is with the School of Information Science and Engineering, Southeast University, Nanjing, China. Email: hehengtao@seu.edu.cn.

	The source code and trained checkpoints are available at \url{https://github.com/weiiguo/dl-distribution-detection}.
}
}

\maketitle

\begin{abstract}
Distributed detection plays a fundamental role in communication-efficient collaborative inference over large-scale networks, where spatially distributed nodes must cooperate using severely quantized messages due to bandwidth constraints. While low-bit quantization is essential for reducing communication overhead, it can significantly degrade detection performance, particularly when prior knowledge of observation noise is unavailable. To address this challenge, we propose a \textit{model-driven deep learning} framework for distributed detection with low-bit quantization. \textcolor{black}{For fixed local quantizers, we first derive the exact finite-sensor maximum a posteriori (MAP) detection error and prove that, \textcolor{black}{under conditionally independent and identically distributed (i.i.d.) observations, independent local randomizations, and a common quantization-level constraint, identical quantizers preserve the supremum of the average network Chernoff information}. Guided by these results, we develop a binary probabilistic quantizer with a trainable neural probability controller and design a tree-structured probabilistic quantizer for scalable multi-bit quantization. The exact finite-sensor MAP error provides the training objective for the binary quantizer, while we derive a computationally tractable Bhattacharyya surrogate objective for training the multi-bit quantizer. For both cases, we derive analytical MAP fusion rules based on the class-conditional probability mass functions (PMFs) of the quantized messages. When the observation statistics are unknown, these PMFs are estimated from held-out labeled data, yielding plug-in MAP detectors. Simulation results show that the proposed method approaches centralized-oracle performance at higher quantization resolutions and remains effective under changes in network size and test-time signal amplitude.}
\end{abstract}

\vspace{-1em}

\section{Introduction}\label{sec:intro}
The increasing prevalence of large-scale interconnected intelligent systems, such as the massive Internet of Things (IoT) \cite{guo2021enabling}, smart grids \cite{galli2011grid}, and multi-agent robotic networks \cite{afrin2021resource}, has generated a pressing demand for efficient distributed inference and learning \cite{mao2018deep,liang2024communication}. In these systems, spatially distributed sensors or agents must collaborate to perform inference and decision-making tasks based on locally observed data, while operating under stringent communication constraints imposed by limited bandwidth and energy resources \cite{chong2003sensor}. Among various inference primitives, distributed detection plays a fundamental role, serving as a building block for collaborative sensing, learning, and decision-making \cite{meng2011collaborative, gururaj2023collaborative}. However, although aggressive quantization of local observations is needed to reduce communication overhead, it inevitably degrades detection performance and poses significant challenges to the design of communication-efficient distributed detection algorithms. Addressing this trade-off between inference accuracy and communication efficiency has therefore become a critical problem in the evolution towards edge intelligence and large-scale distributed learning systems \cite{letaief2019roadmap,letaief2021edge}.

Distributed detection and decentralized hypothesis testing have been extensively studied as fundamental inference primitives in networked systems \cite{tenney1981detection, viswanathan1997distributed, zhang2002optimal}. Early foundational works established that, for independent observations, likelihood ratio tests (LRTs) are optimal under both Bayesian and Neyman-Pearson criteria for a broad class of network architectures, including parallel and serial configurations \cite{luo2017serial}. Subsequent studies investigated the impact of network topology, local decision rules, and fusion strategies on detection performance. In particular, Chamberland and Veeravalli \cite{chamberland2003decentralized} analyzed decentralized detection with independent sensors and characterized asymptotic performance behavior as the number of agents increases. More recently, theoretical efforts have focused on quantifying the performance loss induced by decentralization, identical local decisions, and quantization constraints. Notably, G{\"u}l \cite{gul2025theoretical} derived worst-case bounds on detection performance loss attributable to the network structure, decision rule constraints, and quantization, providing valuable insights into the fundamental limits of decentralized hypothesis testing. While these works offer strong theoretical characterizations, they primarily aim at performance analysis and bounding, rather than the design of scalable, communication-efficient detection algorithms.

Quantization is a key enabler of distributed detection under communication constraints, as local observations or sufficient statistics must be compressed prior to transmission. Classical approaches often considered binary or low-bit decision forwarding, which significantly reduces communication overhead but may incur substantial performance degradation relative to centralized detection, particularly in finite-sensor regimes \cite{tsitsiklis1988decentralized}. To mitigate this loss, multi-bit quantization of local likelihood ratios or sufficient statistics has been widely investigated. Information-theoretic criteria such as Chernoff information and the deflection criterion have emerged as powerful tools for quantizer design, as they provide tractable surrogates for detection error probability and capture the exponential decay rate of error \cite{chen2012modeling}. Building on this line of work, Altay and Deli\c{c} \cite{altay2016optimal} studied optimal multi-bit quantization intervals for distributed detection using Chernoff information and the deflection criterion. They showed that, for independent and identically distributed (i.i.d.) sensor observations, locally optimal quantization, designed independently at each sensor, achieves global optimality at the fusion center. In the more recent works \cite{gul2021fast,gul2021scalable}, the authors proposed a Gaussian approximation method for fast and scalable multilevel quantization and fusion center (FC) design with both i.i.d. and non-i.i.d. sensors. \textcolor{black}{However, current state-of-the-art approaches for distributed detection generally assume exact knowledge of the hypothesis priors and observation distributions, which may be unavailable in practice.}

To address the above challenges, specifically the reliance on perfect statistical knowledge and the computational intractability of optimal high-dimensional design, this paper proposes a model-driven deep learning (DL) method \cite{he2020model} for communication-constrained networks with low-bit quantization. 
Unlike purely data-driven black-box deep learning approaches, our proposed framework \textcolor{black}{combines detection-theoretic structure with sample-driven scalability under stringent communication constraints, retaining the rigor of MAP fusion while leveraging the adaptability of DL to handle unknown observation statistics}. 
\textcolor{black}{This framework builds upon our earlier conference study \cite{guo2024model}, which was limited to binary quantization. The present work extends that study by developing the multi-bit theory and a tree-structured probabilistic quantizer with a tractable training objective, establishing an analytical plug-in MAP fusion framework for both binary and multi-bit messages, and providing substantially expanded experimental validation.} The main contributions of the resulting framework are summarized as follows:
\begin{enumerate}
	\item\textcolor{black}{\textbf{Finite-Sensor MAP Analysis and Quantizer-Deployment Structure:} We derive the exact finite-sensor detection error under the maximum a posteriori (MAP) criterion for fixed local quantizers. \textcolor{black}{For conditionally i.i.d. observations, independent local randomizations, and a common quantization-level constraint, we show that identical quantizers preserve the supremum of the average network Chernoff information.} Under the adopted identical-quantizer architecture, we further show that the binary message average and the multi-bit message histogram preserve the minimum MAP error achieved using all individual messages, thereby enabling fixed-dimensional fusion statistics.}
	\item\textbf{Scalable Probabilistic Quantizer and Analytical Detector Design:} Building on the above theoretical results, we propose a probabilistic quantization framework that supports both binary and multi-bit quantization. For the binary case, we design a probabilistic quantizer combined with a learnable neural probability controller and random dithering. For the multi-bit case, we extend the design into a tree-structured quantizer with sequential random dithering that hierarchically encodes observations into bits. \textcolor{black}{For both cases, the fusion center implements an analytical MAP rule using the class-conditional probability mass functions of the quantized outputs.}
	\item\textcolor{black}{\textbf{Model-Driven Training and Plug-In MAP Fusion:} We translate the above detection-theoretic analysis into resolution-specific model-driven training objectives. For binary quantization, the derived exact finite-sensor MAP error is directly optimized. For multi-bit quantization, whose exact MAP objective requires combinatorial enumeration of message histograms, we derive a computationally tractable Bhattacharyya surrogate that enables scalable controller training. After training, the induced class-conditional message PMFs are estimated from independent held-out labeled data and substituted into the derived analytical MAP rules, yielding plug-in fusion without requiring knowledge of the continuous observation distributions.}
\end{enumerate}

The remainder of this paper is organized as follows: Section \ref{sec:model_problem} introduces the system model and presents the problem formulation. Section \ref{sec:quantizer_FC_design} presents the model-driven DL framework for binary quantization. Section \ref{sec:multi_bit} extends the design to the multi-bit scenario. \textcolor{black}{Section \ref{sec:training_pmf_estimation} describes controller training and post-training PMF estimation.} Section \ref{sec:simulation} provides the numerical results and analysis. Finally, Section \ref{sec:conclusion} concludes this paper.\vspace{-1em}

\section{System Model and Problem Formulation}\label{sec:model_problem}
\subsection{System Model}\label{subsec:system_model}
\begin{figure}[!htbp]
	\centering
	\includegraphics[width=0.45\textwidth]{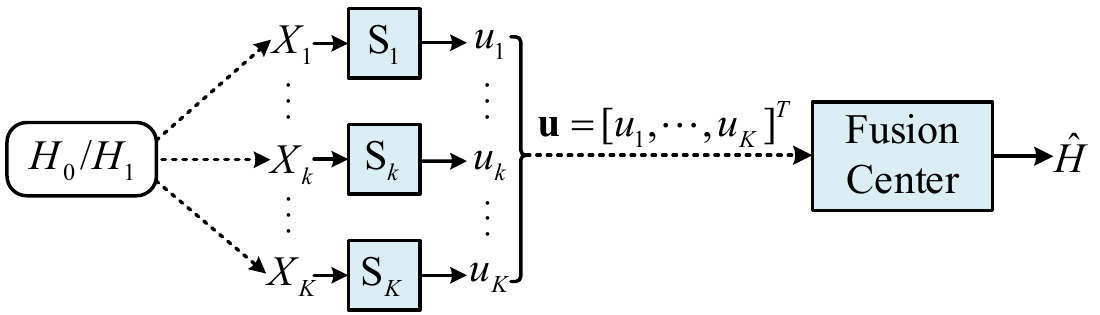}
	\caption{System Model.}\vspace{-0.3cm}
	\label{fig:system}
\end{figure}

As shown in Fig.~\ref{fig:system}, we consider distributed detection in a communication-constrained network, where the FC aims to infer the underlying state of nature based on observations received from $K$ distributed sensors, denoted by $\mathrm{S}_k$, $k=1,\ldots,K$. Specifically, we formulate this as a distributed hypothesis testing problem, where the true hypothesis $H$ belongs to the binary set $\{H_0,H_1\}$ with prior probabilities $P(H=H_0)=\pi_0$ and $P(H=H_1)=\pi_1$. Each sensor $\mathrm{S}_k$, $k=1,\ldots,K$, independently observes $H$ and obtains a local observation $X_k$, which is a noisy version of $H$. 

We consider the scenario in which the observation noise at all sensors is i.i.d.; thus, the local observations are conditionally i.i.d. given $H$, i.e., 
\begin{equation}\label{eq:local_noise_distribution}
	\begin{aligned}
		&f_{X_1, \ldots, X_K \mid H}\left(x_1, \ldots, x_K \mid H_i\right)\\
		&=\prod_{k=1}^K f_{X_k \mid H}\left(x_k \mid H_i\right)=\prod_{k=1}^K f_{X \mid H}\left(x_k \mid H_i\right), i=0,1,
	\end{aligned}
\end{equation} 
where $x_1,\ldots,x_K\in\mathbb{R}$, $f_{X_1,\ldots,X_K\mid H}(\cdot\mid\cdot)$ is the conditional joint PDF of all observations, and $f_{X\mid H}(\cdot\mid\cdot)=f_{X_1\mid H}(\cdot\mid\cdot)=\cdots=f_{X_K\mid H}(\cdot\mid\cdot)$ is their common conditional marginal PDF.

\textcolor{black}{We consider a communication-constrained setting in which each sensor transmits only a finite-alphabet representation of its local observation to the FC.
Specifically, sensor $\mathrm{S}_k$, $k=1,\ldots,K$, maps its continuous-valued observation $X_k$ to a discrete message $u_k\in\mathcal{U}$, where
\begin{equation}
    \mathcal{U}=\{u_0,u_1,\ldots,u_{L-1}\}, \qquad L=2^M.
\end{equation}
Here, $M$ is a prescribed quantization resolution common to all sensors.
Thus, each message can be represented by an $M$-bit index, and the total sensor-to-FC payload is $KM$ bits per detection instance, instead of the original real-valued observations.} 
Let the conditional distribution $p(u_k|X_k)$ describe the probabilistic quantization process at sensor $k$, specifying the distribution of the quantized message $u_k$ given the local observation $X_k$.
Subsequently, each sensor $\mathrm{S}_k$, $k=1,\ldots,K$, transmits $u_k$ to the FC through an error-free channel\footnote{\textcolor{black}{The communication limitation is the prescribed $M$-bit payload per sensor, which necessitates lossy quantization, whereas the error-free assumption concerns only the reliable delivery of the resulting message.}}, and the FC aggregates the quantized data $\mathbf{u}=[u_1,\ldots,u_K]^T$ from all $K$ sensors to generate the decision $\hat{H}$.

\subsection{Problem Formulation}\label{subsec:problem}
To evaluate the detection performance, we utilize the detection error probability $P_e(H,\mathbf{u},\hat{H})$ as the cost function, i.e.,
\begin{equation}\label{eq:detection_error}
	\begin{aligned}
		P_e(H,\mathbf{u},\hat{H})=&P\left(\hat{H}\neq H\right)\\
		=&\pi_{0}\sum_{\mathbf{u}}P\left(\hat{H}(\mathbf{u})=H_{1}|\mathbf{u}\right)P\left(\mathbf{u}|H_{0}\right)\\
		&+\pi_{1}\sum_{\mathbf{u}}P\left(\hat{H}(\mathbf{u})=H_{0}|\mathbf{u}\right)P(\mathbf{u}|H_{1}).
	\end{aligned}
\end{equation}

Our goal is to find the optimal probabilistic quantizers $\{p(u_k|X_k)\}_{k=1}^K$ and detector $\hat{H}(\mathbf{u})$ to minimize $P_e(H,\mathbf{u},\hat{H})$, and the corresponding optimization problem is formulated as
\begin{equation}\label{eq:opt_problem}
	\underset{\left\{p(u_k|X_k)\right\}_{k=1}^{K},\,
		\hat{H}(\cdot)}{\min}\
	P_e(H,\mathbf{u},\hat{H}).
\end{equation}

\textcolor{black}{
A classical result in Bayesian hypothesis testing states that, for any fixed set of quantizers, the MAP decision rule minimizes the detection error probability
\cite{cover2006elements}. We therefore employ the MAP detector at the FC,
given by
\begin{equation}\label{eq:MAP_rule}
    \hat{H}_{\mathrm{MAP}}(\mathbf{u})=
    \begin{cases}
        H_0, &
        \pi_0p(\mathbf{u}|H_0)>
        \pi_1p(\mathbf{u}|H_1),\\
        H_1, &
        \pi_0p(\mathbf{u}|H_0)\leq
        \pi_1p(\mathbf{u}|H_1).
    \end{cases}
\end{equation}
}
\textcolor{black}{
Substituting \eqref{eq:MAP_rule} into \eqref{eq:detection_error}, we note
that, for each realization $\mathbf{u}$, deciding $H_0$ incurs the error
term $\pi_1p(\mathbf{u}|H_1)$, whereas deciding $H_1$ incurs
$\pi_0p(\mathbf{u}|H_0)$. The MAP detector selects the decision associated
with the smaller error term. Hence, its detection error probability is
\begin{equation}\label{eq:MAPDEP}
    \begin{aligned}
        P_E(H,\mathbf{u})
        &\triangleq
        P_e(H,\mathbf{u},\hat{H}_{\mathrm{MAP}})\\
        &=\sum_{\mathbf{u}}
        \min\left\{
        \pi_0p(\mathbf{u}|H_0),
        \pi_1p(\mathbf{u}|H_1)
        \right\}.
    \end{aligned}
\end{equation}
We refer to $P_E(H,\mathbf{u})$ as the MAP detection error probability
(MAPDEP).
}
\textcolor{black}{
\begin{remark}\label{rem:MAPDEP_metric}
	Equation \eqref{eq:MAPDEP} shows that, after adopting the optimal MAP detector, the minimum detection error probability depends on the local quantizers only through the induced conditional distributions $p(\mathbf{u}|H_0)$ and $p(\mathbf{u}|H_1)$. Therefore, the detector need not be optimized separately, and $P_E(H,\mathbf{u})$ provides an exact finite-sensor objective for quantizer design. 
\end{remark}
}
\textcolor{black}{
Although \eqref{eq:MAPDEP} gives the exact minimum detection error probability for a finite number of sensors, directly optimizing it over possibly heterogeneous local quantizers remains difficult. We therefore consider the average network Chernoff information \cite{cover2006elements}, defined as
\begin{equation}\label{eq:network_Chernoff_information}
    C(H,\mathbf{u})
    \triangleq
    -\frac{1}{K}
    \min_{0\leq\lambda\leq 1}
    \log\left[
        \sum_{\mathbf{u}}
        p(\mathbf{u}|H_0)^\lambda
        p(\mathbf{u}|H_1)^{1-\lambda}
    \right].
\end{equation}
For conditionally i.i.d. quantized messages, Chernoff information characterizes the asymptotic error exponent of the optimal Bayesian detector, which provides a tractable information-theoretic criterion for studying the structure of the quantizer deployment. The following lemma establishes the optimal deployment structure under this criterion.
}
\textcolor{black}{
\begin{lemma}\label{lem:glob_opt_identical_quantizer}
	Suppose that the local observations are conditionally i.i.d. given $H$, and that all sensors are subject to the same $L$-level quantization constraint. \textcolor{black}{Assume that the local quantization randomizations are mutually independent across sensors and independent of $(H,X_1,\ldots,X_K)$. Then restricting all sensors to an identical quantizer does not change the supremum of the average network Chernoff information over all possibly heterogeneous local quantizers.} Specifically,
	\begin{equation}\label{eq:max_Chernoff}
		\begin{aligned}
			&\textcolor{black}{\sup_{\{p(u_k|X_k)\}_{k=1}^{K}}}C(H,\mathbf{u})=\\
			&\textcolor{black}{\sup_{p(u|X)}}\left\{-\min_{0\leq\lambda\leq 1}\log\left[\sum_{u=0}^{L-1}p(u|H_{0})^{\lambda}p(u|H_{1})^{1-\lambda}\right]\right\},
		\end{aligned}
	\end{equation}  
	where $p(u|H_i)=\mathbb{E}_{X|H_i}[p(u|X)],\ i=0,1$. In particular, if $p(u|X)^*$ attains the single-sensor \textcolor{black}{supremum} on the right-hand side of \eqref{eq:max_Chernoff}, deploying $p(u|X)^*$ at every sensor is globally optimal for maximizing the average network Chernoff information. 
\end{lemma}
\begin{IEEEproof}
		The proof is provided in Appendix~\ref{append:proof_identical_quantizer}. 
\end{IEEEproof}
}
\textcolor{black}{
\textcolor{black}{Under the assumptions of Lemma~\ref{lem:glob_opt_identical_quantizer}, restricting all sensors to an identical
quantizer incurs no loss in the supremum of the average network Chernoff information.
Guided by this structural result, we adopt a shared $L$-level quantizer and
optimize the finite-sensor MAP error within this architecture. The resulting
restricted optimization problem is}
\begin{equation}\label{eq:opt_problem_quantizer}
    \underset{p(u|X)}{\min}\
    P_E(H,\mathbf{u}).
\end{equation}
}

\textcolor{black}{This formulation provides the basis for the model-driven design developed in the following sections.}

\section{Model-Driven DL for Binary Quantization and Detector Design}\label{sec:quantizer_FC_design}
In this section, we translate the theoretical results established in Section \ref{sec:model_problem} into a concrete and learnable system design. In particular, we consider the binary quantizer and detector design, and develop a model-driven DL framework \textcolor{black}{guided by the derived MAPDEP and Chernoff structural result}.

\subsection{Binary Quantizer Design}\label{subsec:binary_quantizer}
First, we consider the binary quantization scheme. \textcolor{black}{Guided by Lemma \ref{lem:glob_opt_identical_quantizer}, we restrict attention to an identical binary quantizer across sensors}, and the following proposition gives a closed-form expression of MAPDEP in this case.
\begin{proposition}\label{prop:MAPDEP_binary_quantize}
	If the binary quantized data $u_1,\ldots,u_K$ from all sensors are conditionally i.i.d. given $H$, then the MAPDEP under binary quantization is given by
	\begin{equation}\label{eq:MAPDEP_Binary_Quantization}
		\begin{aligned}
			P_E^{\text{Binary}}\triangleq\sum_{k=0}^{K}C_{K}^{k}\min&\left\{\pi_0(\gamma(H_0))^k(1-\gamma(H_0))^{K-k},\right.\\
			&\left.\pi_1(\gamma(H_1))^k(1-\gamma(H_1))^{K-k}\right\},
		\end{aligned}
	\end{equation}
	where 
	\begin{equation}\label{eq:binary_output_probability}
		\gamma(H)=P(u=1|H)=\mathbb{E}_{X|H}[p(u=1|X)],\ H\in\{H_0,H_1\},
	\end{equation}
		denotes the conditional probability that a quantized message equals `1' given $H$, and $C_K^k$ is the binomial coefficient.
\end{proposition}
\begin{IEEEproof}
		Please see Appendix~\ref{append:proof_MAPDEP_binary_quantize}.
\end{IEEEproof}

According to Proposition \ref{prop:MAPDEP_binary_quantize}, when identical quantizers are employed across sensors, the MAPDEP depends on the quantizer only through the conditional probabilities \eqref{eq:binary_output_probability}. This observation implies that, from the optimization perspective, the quantizer design problem can be equivalently viewed as learning the conditional output probability $P(u=1|X)$, rather than directly determining a deterministic decision rule. This formulation naturally motivates the use of probabilistic quantizers, which provide a continuous and differentiable representation of the quantization behavior.

\begin{figure}[!htbp]
	\centering
	\includegraphics[width=0.45\textwidth]{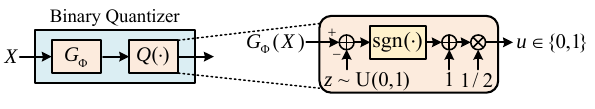}
	\caption{Proposed Binary Quantizer.}
	\label{fig:quantizer}
\end{figure}

However, while probabilistic quantization is convenient for analysis and learning, practical sensor networks require the transmission of hard binary symbols. To bridge this gap between the probabilistic model and practical implementation, we adopt a binary probabilistic quantizer with random dithering \cite{gray1993dithered}, as shown in Fig.~\ref{fig:quantizer}. Specifically, the local observation $X$ is first sent to a probability controller $G(\cdot): \mathbb{R}\rightarrow [0,1]$, and then the output $G(X)$ is fed into a quantization function $Q(\cdot):[0,1]\rightarrow\{0,1\}$ to generate a random binary symbol $u$, given by
\begin{equation}\label{eq:quantization_func}
	u=Q(G(X))=\frac{1+\text{sgn}(G(X)-z)}{2}\in\{0,1\}.
\end{equation} 
Here, $z\sim U(0,1)$ is a standard uniform random variable serving as dithering noise, and $\text{sgn}(\cdot)$ is the sign function. As shown in \cite{he2024joint}, this quantizer ensures that $P(u=1|X)=G(X)$ and $P(u=0|X)=1-G(X)$. 

Our goal is to find the optimal binary quantizer to minimize $P_E^{\text{Binary}}$ in \eqref{eq:MAPDEP_Binary_Quantization}, and the key is to find the optimal probability controller $G(\cdot)$. \textcolor{black}{However, $P_E^{\text{Binary}}$ depends on class-conditional probabilities that may be difficult to obtain.} This motivates us to use the model-driven DL method \cite{he2020model} to train a deep neural network (DNN), parameterized by $\Phi\in\mathbb{R}^d$, to approximate the optimal probability controller $G^*(\cdot)$.

Since the MAPDEP $P_E^{\text{Binary}}$ provides a reliable performance metric for the binary quantizer, it can be directly used as the loss function to optimize the parameters $\Phi$. Let $G_{\Phi}(\cdot)$ denote the neural probability controller parameterized by $\Phi$. The loss function for training $\Phi$ is
\begin{equation}\label{eq:loss_func_quantizer}
	\begin{aligned}
		\mathcal{L}_{\Phi}=\sum_{k=0}^{K}C_{K}^{k}\min&\{\pi_0(\gamma_{\Phi}(H_0))^k(1-\gamma_{\Phi}(H_0))^{K-k},\\
		&\pi_1(\gamma_{\Phi}(H_1))^k(1-\gamma_{\Phi}(H_1))^{K-k}\},
	\end{aligned}
\end{equation}
with $\gamma_{\Phi}(H_i)\triangleq \mathbb{E}\!\left[G_{\Phi}(X)\mid H_i\right],\ i=0,1$,
where the expectation is taken with respect to the conditional distribution of $X$ given $H_i$.

\subsection{\textcolor{black}{Analytical MAP Detector Design}}\label{subsec:binary_detector}
After obtaining the DNN-based binary quantizer, we further need to find the optimal detector at the FC to minimize $P_e(H,\mathbf{u},\hat{H})$ in the original problem \eqref{eq:opt_problem}. As shown in Section \ref{subsec:system_model}, the FC aggregates the quantized data $\mathbf{u}\in\{0,1\}^K$ from all sensors to generate the final decision $\hat{H}$. Based on the MAP decision rule, the following theorem establishes the equivalence between using all quantized data and using their average as the detector input.
\begin{theorem}\label{th:opt_mean_fusion}
	If the binary quantized data $u_k$, $k=1,\ldots,K$, from all sensors are conditionally i.i.d. given $H$, then detection using the quantized data $\mathbf{u}=[u_1,\dots,u_K]^T$ or their average $\bar{u} = \frac{1}{K}\sum_{k=1}^{K} u_k$ achieves the same minimum detection error probability, i.e.,
	\begin{equation}\label{eq:mean_fusion_MAPDEP}
		\begin{aligned}
			P_{E}(H,\mathbf{u})=&P_E^{\text{Binary}}\\
			=&P_{E}(H,\bar{u})\triangleq\sum_{\bar{u}}\min\left\{\pi_0p(\bar{u}|H_0),\pi_1p(\bar{u}|H_1)\right\},
		\end{aligned}
	\end{equation}
		where $P_{E}(H,\bar{u})$ is the minimum detection error probability achieved by using $\bar{u}$ to detect $H$.
\end{theorem}
\begin{IEEEproof}
		Please see Appendix~\ref{append:proof_mean_fusion}.
\end{IEEEproof}

\textcolor{black}{According to Theorem \ref{th:opt_mean_fusion}, we can use the average $\bar{u}$ instead of all the quantized data $\mathbf{u}$ as the input for the detector. Then, the MAP decision for the detection of $H$ is given by
\begin{equation}\label{eq:MAP_detector_average}
	\hat{H}_{\text{MAP}}(\bar{u})=\arg\max_{H}\ p(H|\bar{u}).
\end{equation}
\textcolor{black}{However, \eqref{eq:MAP_detector_average} is not directly implementable when the posterior $p(H\mid\bar{u})$ is unavailable. The remaining challenge is to express it in terms of the low-dimensional class-conditional probabilities induced by the quantizer, which can subsequently be estimated from labeled data.}
Let $s\triangleq\sum_{k=1}^{K}u_k=K\bar{u}$ denote the number of sensors transmitting $1$. Under the conditionally i.i.d. assumption, the individual quantized messages are Bernoulli conditioned on each hypothesis. Therefore, for $i=0,1$, the conditional PMF of $s$ is
\begin{equation}\label{eq:binary_count_likelihood}
	\begin{aligned}
	p_{\Phi}&(s\mid H_i)=\\
	&C_K^s\left(\gamma_{\Phi}(H_i)\right)^s\left(1-\gamma_{\Phi}(H_i)\right)^{K-s},\ s=0,1,\ldots,K,
	\end{aligned}
\end{equation}
where $\gamma_{\Phi}(H_i)=p_{\Phi}(u_k=1\mid H_i)$ is the class-conditional probability of the binary quantized message. According to Bayes' rule, the posterior probability of $H_i$ given $s$ is
\begin{equation}\label{eq:binary_MAP_posterior}
	p_{\Phi}(H_i\mid s)
	=
	\frac{
		\pi_i
		\left(\gamma_{\Phi}(H_i)\right)^s
		\left(1-\gamma_{\Phi}(H_i)\right)^{K-s}
	}{
		\displaystyle
		\sum_{j=0}^{1}
		\pi_j
		\left(\gamma_{\Phi}(H_j)\right)^s
		\left(1-\gamma_{\Phi}(H_j)\right)^{K-s}
	}.
\end{equation}
Note that the common binomial coefficient $C_K^s$ in \eqref{eq:binary_count_likelihood} cancels when computing the posterior probability.}

\textcolor{black}{
Accordingly, the MAP detector at the FC can be equivalently implemented using
the following log-posterior odds as
\begin{equation}\label{eq:binary_analytical_MAP_LLR}
	\begin{aligned}
		\Lambda_{\Phi}(s)
		\triangleq&
		\log
		\frac{\pi_1p_{\Phi}(s\mid H_1)}
		     {\pi_0p_{\Phi}(s\mid H_0)}
		\\
		=&
		\log\frac{\pi_1}{\pi_0}
		+s\log
		\frac{\gamma_{\Phi}(H_1)}
		     {\gamma_{\Phi}(H_0)}
		+(K-s)\log
		\frac{1-\gamma_{\Phi}(H_1)}
		     {1-\gamma_{\Phi}(H_0)}.
	\end{aligned}
\end{equation}
The corresponding MAP decision rule is given by
\begin{equation}\label{eq:binary_analytical_MAP_rule}
	\hat{H}_{\mathrm{MAP}}(s)
	=
	\begin{cases}
		H_1, & \Lambda_{\Phi}(s)\geq 0,\\
		H_0, & \Lambda_{\Phi}(s)<0.
	\end{cases}
\end{equation}
Since $s=K\bar{u}$, \eqref{eq:binary_analytical_MAP_rule} can be directly evaluated by using the average quantized message $\bar{u}$.
The rule in \eqref{eq:binary_analytical_MAP_rule} is the analytical MAP
detector for the class-conditional Bernoulli probabilities induced by a given
quantizer. The estimation of these probabilities when the observation
statistics are unknown is separated from the detector derivation and presented
in Section \ref{sec:training_pmf_estimation}.
}

\section{Model-Driven DL for Multi-Bit Quantization and Detector Design}\label{sec:multi_bit}
In this section, we extend the binary quantization framework to a multi-bit scenario. Specifically, by employing an $M$-bit quantization scheme, the local observation $X$ at each sensor is quantized into one of $L=2^M$ discrete quantization levels, denoted by the set $\mathcal{U} = \{u_0, u_1, \dots, u_{L-1}\}$. To facilitate efficient transmission in communication-constrained networks, each quantization level $u_l$ is represented by an $M$-bit binary codeword $\mathbf{b}=[b_1,b_2,\ldots,b_M]^T \in \{0, 1\}^M$ corresponding to its index $l$, and the bijective mapping between $l$ and $\mathbf{b}$ is given by
\begin{equation}\label{eq:index_bit_mapping}
    l = \sum_{m=1}^{M} b_m 2^{M-m}.
\end{equation}

\subsection{Tree-based Quantizer Design}
In this section, we introduce a hierarchical method for generating the binary codeword $\mathbf{b}$. We model the $M$-bit quantization process as a traversal of a binary decision tree of depth $M$. The tree consists of an internal-node set $\mathcal{V}$ and a leaf-node set $\mathcal{U}$. The root is indexed by $v=0$, and for any internal node $v$, its left and right children, corresponding to bits 0 and 1, are indexed by $2v+1$ and $2v+2$, respectively. Therefore, the tree has $|\mathcal{V}|=2^M-1$ internal nodes.

\subsubsection{Tree Probability Controller}
Similar to the binary case, we employ a DNN-based probability controller, parameterized by $\Phi$, to map the local observation $X$ to the quantization probabilities. However, instead of a single scalar, the controller $G_{\Phi}(\cdot)$ in this multi-bit case outputs the conditional branching probabilities for all internal nodes of the complete binary decision tree:
\begin{equation}\label{eq:branching_prob}
	\bm{\gamma}(X)=\left[\gamma_0(X),\gamma_1(X),\ldots,\gamma_{2^M-2}(X)\right],
\end{equation}
where $\gamma_{v}(X)\in[0,1]$ denotes the conditional probability of taking the right child, i.e., generating bit 1, given that the traversal has reached internal node $v$. Mathematically,
\begin{equation}
	\gamma_{v}(X)\triangleq P\left(\text{corresponding bit}=1\mid \text{current node}=v, X\right).
\end{equation}
Correspondingly, $1-\gamma_{v}(X)$ represents the probability of branching to the left, i.e., generating bit 0.

\subsubsection{Induced Joint Distribution over Codewords}
The tree structure induces a joint distribution over all $M$-bit codewords. 
For any quantization level $u_l\in\mathcal{U}$, let 
$\mathbf{b}^{(l)}=[b_1^{(l)},b_2^{(l)},\ldots,b_M^{(l)}]^T$ 
denote its corresponding binary codeword. This codeword uniquely determines a path from the root to a leaf in the binary tree. 
Using a level-order indexing with the root node indexed as $v_0^{(l)}=0$, 
the sequence of internal nodes visited along the path is given by
$\{v_0^{(l)},v_1^{(l)},\ldots,v_{M-1}^{(l)}\}$, where
\begin{equation}\label{eq:traversal_rule}
	v_{m}^{(l)} = 2v_{m-1}^{(l)} + b_{m}^{(l)}+1, \quad m=1,\dots,M,
\end{equation}
and $b_m^{(l)}=0$ and $b_m^{(l)}=1$ correspond to the left and right child, respectively. Here, $v_M^{(l)}$ is the leaf node reached after $M$ steps.

Conditioned on the observation $X$, the probability of generating codeword 
$\mathbf b^{(l)}$ and thus selecting level $u_l$ is given by
\begin{small}
	\begin{equation}\label{eq:codeword_prob}
		\begin{aligned}
			P(u=u_l\mid X)\triangleq & P(\mathbf b=\mathbf b^{(l)}\mid X)\\
			\triangleq & \prod_{m=1}^{M}\left(\gamma_{v_{m-1}^{(l)}}(X)\right)^{b_m^{(l)}}\left(1-\gamma_{v_{m-1}^{(l)}}(X)\right)^{1-b_m^{(l)}}.
		\end{aligned}
	\end{equation}
\end{small}

Let $\mathbf q(X)\in[0,1]^L$ denote the induced probability distribution over all
$L$ quantization levels, where the $l$-th entry is
$[\mathbf q(X)]_l = P(u=u_l\mid X)$.
This distribution can be efficiently computed via a hierarchical traversal
of the tree without explicit sampling.

\subsubsection{Codeword Generation with Sequential Dithering}\label{subsubsec:sequential_dithering}
Based on the above tree structure, the quantizer generates the binary codeword $\mathbf{b}$ in a sequential manner by traversing the binary tree from the root to a leaf. The procedure proceeds for $M$ steps, corresponding to the $M$ bits of the
codeword.

Specifically, the traversal starts at the root node indexed as $v_0=0$.
At the $m$-th step, suppose the traversal is currently at internal node $v_{m-1}$.
An independent dithering variable $z_m\sim U(0,1)$ is drawn, and the $m$-th bit is generated as
\begin{equation}\label{eq:tree_dither}
	b_m = \frac{1+\text{sgn}(\gamma_{v_{m-1}}(X)-z_m)}{2},
\end{equation}
and the next visited node is then updated according to \eqref{eq:traversal_rule}. After $M$ steps, the traversal reaches a leaf node, yielding the output
codeword $\mathbf b=[b_1,\dots,b_M]^T$ and the corresponding quantization level
$u_l$. 

In our proposed tree-based quantizer, the probability controller $G_\Phi(\cdot)$
outputs the node-wise branching probabilities $\bm{\gamma}(X)=\{\gamma_v(X)\}_{v\in\mathcal V}$, as in \eqref{eq:branching_prob}. These probabilities deterministically induce an $L$-ary distribution
$\mathbf q(X)\in[0,1]^L$ over quantization levels via the tree structure, where
$[\mathbf q(X)]_l=P(u=u_l\mid X)$ is computed by hierarchical traversal without explicit sampling. Meanwhile, in practical deployment, the quantizer must produce a \emph{hard} quantization output by sequentially generating bits using dithering noise and traversing the tree according to \eqref{eq:traversal_rule}.

The following proposition establishes a key property of the proposed tree-based quantizer: the sequential dithering procedure realizes exactly the \emph{model-induced} distribution defined in \eqref{eq:codeword_prob}. This property allows us to avoid stochastic sampling during training while ensuring consistency with the deployment-stage quantization.
\begin{proposition}\label{prop:tree_distribution}
	Given an observation $X$, suppose the binary codeword $\mathbf b=[b_1,\dots,b_M]^T$ is generated according to the sequential dithering procedure. Denote the induced distribution of this random output by $P_{\mathrm{SD}}(\cdot\mid X)$. Then, for any level $u_l\in\mathcal U$ with associated codeword $\mathbf b^{(l)}=[b_1^{(l)},\dots,b_M^{(l)}]^T$ and induced internal-node sequence $\{v_0^{(l)},v_1^{(l)},\dots,v_{M-1}^{(l)}\}$ defined by \eqref{eq:traversal_rule}, we have
	\begin{equation}\label{eq:prop_level}
		P_{\mathrm{SD}}(u=u_l\mid X) = [\mathbf q(X)]_l,
	\end{equation}
	where $[\mathbf q(X)]_l$ is the model-induced probability defined in \eqref{eq:codeword_prob}.
\end{proposition}

\begin{IEEEproof}
	The proof is provided in Appendix~\ref{append:proof_tree_distribution}.
\end{IEEEproof}

Next, we characterize the minimum achievable MAP detection error probability under the proposed tree-structured multi-bit quantizer. Assuming identical quantizers are deployed at all sensors, we define the conditional probability of the quantized output $u$ taking the quantization level $u_l$ under hypothesis $H_i$ as
\begin{equation}
	\gamma_l(H_i)\triangleq P(u=u_l \mid H_i) = \mathbb{E}_{X\mid H_i}\left[P(u=u_l \mid X)\right],\ \forall l,
\end{equation}
where $\sum_{l=0}^{L-1}\gamma_l(H_i)=1$ for $i\in\{0,1\}$. Let $\mathbf{n}=[n_0,n_1,\ldots,n_{L-1}]^T$ denote the count vector, where $n_l$ represents the number of sensors that output quantization level $u_l$. Since there are $K$ sensors in total, the elements of $\mathbf{n}$ must satisfy $\sum_{l=0}^{L-1}n_l=K$. Let $\mathcal{N}$ denote the set of all valid count vectors:
\begin{equation}
	\mathcal{N}=\left\{\mathbf{n}\in\mathbb{Z}_{\geq 0}^L\middle| \sum_{l=0}^{L-1}n_l=K\right\}.
\end{equation}
Given that local observations are conditionally i.i.d., the quantized outputs from the $K$ sensors follow a multinomial distribution. Consequently, the probability of observing a specific count vector $\mathbf{n}$ given hypothesis $H_i$ is
\begin{equation}\label{eq:likelihood_multi}
	p(\mathbf{n}\mid H_i) = \frac{K!}{n_0!n_1!\cdots n_{L-1}!}\prod_{l=0}^{L-1}\left(\gamma_l\left(H_i\right)\right)^{n_l}.
\end{equation}
Then, according to the MAP decision rule, for any observed count vector $\mathbf{n}$, the detector minimizes the error probability by choosing the hypothesis with the maximum posterior probability. The minimum achievable detection error probability, denoted by $P_E^{\text{Multi}}$, is obtained by summing the minimum of the joint probabilities over all possible outcomes $\mathbf{n}\in\mathcal{N}$, as follows:
\begin{small}
	\begin{equation}\label{eq:multi_bit_MAPDEP}
	\begin{aligned}
		& P_E^{\rm{Multi}}\\
		=&\sum_{\mathbf{n}\in\mathcal{N}}\min\left\{\pi_0p(\mathbf{n}\mid H_0),\pi_1p(\mathbf{n}\mid H_1)\right\}\\
		=&\sum_{\mathbf{n}\in\mathcal{N}}\frac{K!}{\prod_{l=0}^{L-1}n_l!}\min\left\{\pi_0\prod_{l=0}^{L-1}\left(\gamma_l\left(H_0\right)\right)^{n_l},\pi_1\prod_{l=0}^{L-1}\left(\gamma_l\left(H_1\right)\right)^{n_l}\right\}.
	\end{aligned}
\end{equation}
\end{small}

However, unlike the binary case, using the minimum MAPDEP $P_E^{\rm{Multi}}$ in \eqref{eq:multi_bit_MAPDEP} as a training loss is computationally prohibitive due to a severe \emph{combinatorial explosion}. Evaluating \eqref{eq:multi_bit_MAPDEP} requires enumerating all feasible count vectors:
\begin{equation}
	|\mathcal N|
	= \left|\left\{\mathbf{n}\in\mathbb{Z}_{\ge 0}^L:\sum_{l=0}^{L-1}n_l=K\right\}\right|
	= \binom{K+L-1}{L-1}.
\end{equation}
Moreover, each term involves products over $L$ levels and requires computing
$\prod_{l=0}^{L-1} n_l!$ or its logarithm. Therefore, the overall computational
cost scales at least on the order of
\begin{equation}
	\Omega\!\left(|\mathcal N|\cdot L\right)
	= \Omega\!\left(L\binom{K+L-1}{L-1}\right),
\end{equation}
which becomes astronomically large even for moderate $K$ and $M$, as $L=2^M$. For fixed $L$, the number of terms already grows polynomially in $K$
as $\binom{K+L-1}{L-1}=\Omega(K^{L-1})$, whereas increasing $M$ doubles $L$ and
causes the term count to blow up dramatically, making direct evaluation of
\eqref{eq:multi_bit_MAPDEP} intractable in practice.

\textcolor{black}{Therefore, we introduce a tractable surrogate for the exact finite-$K$ MAP detection error. 
For conditionally i.i.d.\ quantized messages, optimizing the Chernoff parameter yields the asymptotic error exponent of the optimal Bayesian detector, but it does not in general give the exact finite-$K$ MAP error probability. 
We therefore use the corresponding Chernoff upper bound as a tractable design criterion, i.e.,
\begin{equation}\label{eq:chernoff_general_multi}
	P_E^{\rm Multi}
	\le
	\pi_0^{\lambda}\pi_1^{1-\lambda}
	\sum_{\mathbf n\in\mathcal N}
	p(\mathbf n\mid H_0)^{\lambda}
	p(\mathbf n\mid H_1)^{1-\lambda}.
\end{equation}}
A particularly convenient and commonly used choice is
$\lambda=1/2$, which reduces \eqref{eq:chernoff_general_multi} to the
Bhattacharyya bound:
\begin{equation}\label{eq:bhatt_bound_multi_step1}
	P_E^{\rm Multi}
	\le
	\sum_{\mathbf n\in\mathcal N}
	\sqrt{\pi_0 p(\mathbf n\mid H_0)\;\pi_1 p(\mathbf n\mid H_1)}.
\end{equation}

Substituting the multinomial likelihood in \eqref{eq:likelihood_multi} into
\eqref{eq:bhatt_bound_multi_step1}, we obtain
\begin{equation}
	P_E^{\rm Multi}
	\le
	\sqrt{\pi_0\pi_1}
	\sum_{\mathbf n\in\mathcal N}
	\frac{K!}{\prod_{l=0}^{L-1}n_l!}
	\prod_{l=0}^{L-1}
	\left(\sqrt{\gamma_l(H_0)\gamma_l(H_1)}\right)^{n_l}.
\end{equation}
Using the multinomial theorem, the summation over $\mathbf n$ admits the closed
form
\begin{equation}\label{eq:multibit_loss}
	P_E^{\rm Multi}
	\le
	\sqrt{\pi_0\pi_1}
	\left(
	\sum_{l=0}^{L-1}\sqrt{\gamma_l(H_0)\gamma_l(H_1)}
	\right)^K,
\end{equation}
which provides a computationally efficient surrogate for training the proposed
tree-structured quantizer.
\textcolor{black}{
\begin{remark}\label{rem:multibit_surrogate_scope}
	Equation \eqref{eq:multibit_loss} provides a computationally tractable Bhattacharyya design objective whose evaluation is linear in the number of quantization levels. Its optimized Chernoff counterpart characterizes the asymptotic Bayesian error exponent, and \eqref{eq:multibit_loss} attains the same exponent when $\lambda^*=1/2$. At finite $K$, it serves as an upper-bound surrogate rather than the exact MAP error.
\end{remark}
}

\subsection{\textcolor{black}{Analytical MAP Detector Design}}\label{subsec:multibit_detector}
When the FC receives $K$ quantized messages, equivalently the collection of binary codewords $\mathbf{B}=[\mathbf{b}_1,\mathbf{b}_2,\ldots,\mathbf{b}_K]^T$, it first transforms them into $K$ one-hot vectors $\mathbf{V}=[\mathbf{v}_1,\mathbf{v}_2,\ldots,\mathbf{v}_K]^T$. The one-hot encoding of $\mathbf{b}_k$ is based on its decimal value and is given by
\begin{equation}\label{eq:one_hot_encoding}
	\mathbf{v}_k= 
		\begin{cases} 
			[1, \mathbf{0}_{L-1}^T]^T, & (\mathbf{b}_k)_{10} = 0 \\
			[\mathbf{0}_{(\mathbf{b}_k)_{10}}^T, 1, \mathbf{0}_{L-1-(\mathbf{b}_k)_{10}}^T]^T, & (\mathbf{b}_k)_{10} \geq 1,
		\end{cases}
\end{equation} 
with $(\mathbf{b}_k)_{10}$ denoting the decimal value of $\mathbf{b}_k$. Then, the FC utilizes the mean one-hot vector $\bar{\mathbf{v}}=(1/K)\sum_{k=1}^{K}\mathbf{v}_k$ to make the MAP decision for $H$ as
\begin{equation}\label{eq:MAP_detection_multi}
	\hat{H}_{\rm MAP}(\bar{\mathbf{v}})=\arg\max_{H} p(H|\bar{\mathbf{v}}).
\end{equation}
Similar to the binary case, we show in the following theorem that the detector at the FC can utilize the mean one-hot vector, thereby reducing its dependence on the number of deployed sensors.
\begin{theorem}\label{th:multibit_mean_fusion}
	If the $M$-bit quantized data from all sensors are conditionally i.i.d. given $H$, then detection using the full quantized data, i.e., the collection of binary codewords $\mathbf{B}=[\mathbf{b}_1,\mathbf{b}_2,\ldots,\mathbf{b}_K]^T$, or their mean one-hot vector $\bar{\mathbf{v}}=\frac{1}{K}\sum_{k=1}^{K}\mathbf{v}_k$ achieves the same minimum detection error probability, i.e.,
	\begin{equation}\label{eq:multi_bit_error_prob}
		\begin{aligned}
			P_E\left(H, \mathbf{B}\right) & = P_E\left(H, \bar{\mathbf{v}}\right)\\
			& = \sum_{\bar{\mathbf{v}}}\min\left\{\pi_0p(\bar{\mathbf{v}} | H_0), \pi_1p(\bar{\mathbf{v}} | H_1)\right\}.
		\end{aligned}
	\end{equation}
\end{theorem}
\begin{IEEEproof}
	The proof is provided in Appendix~\ref{append:th_multibit_mean_fusion}.
\end{IEEEproof}

\textcolor{black}{
We next derive an analytical MAP detector for the multi-bit case. Let
$\mathbf n=\sum_{k=1}^{K}\mathbf v_k=K\bar{\mathbf v}$ be the count vector,
whose $l$-th entry $n_l$ is the number of sensors transmitting the codeword
associated with level $u_l$. From the multinomial likelihood in
\eqref{eq:likelihood_multi}, Bayes' rule gives
\begin{equation}\label{eq:multibit_MAP_posterior}
	p(H_i\mid\mathbf n)
	=
	\frac{
		\displaystyle
		\pi_i\prod_{l=0}^{L-1}\left(\gamma_l(H_i)\right)^{n_l}
	}{
		\displaystyle
		\sum_{j=0}^{1}
		\pi_j\prod_{l=0}^{L-1}\left(\gamma_l(H_j)\right)^{n_l}
	},
	\quad i=0,1.
\end{equation}
The multinomial coefficient in \eqref{eq:likelihood_multi} is common to both
hypotheses and therefore cancels in \eqref{eq:multibit_MAP_posterior}.
}
\textcolor{black}{
Accordingly, the log-posterior odds can be written as
\begin{equation}\label{eq:multibit_analytical_MAP_LLR}
	\begin{aligned}
		\Lambda(\mathbf n)
		&\triangleq
		\log\frac{\pi_1p(\mathbf n\mid H_1)}
		          {\pi_0p(\mathbf n\mid H_0)}\\
		&=
		\log\frac{\pi_1}{\pi_0}
		+\sum_{l=0}^{L-1}
		n_l\log\frac{\gamma_l(H_1)}{\gamma_l(H_0)}\\
		&=
		\log\frac{\pi_1}{\pi_0}
		+K\sum_{l=0}^{L-1}
		[\bar{\mathbf v}]_l
		\log\frac{\gamma_l(H_1)}{\gamma_l(H_0)}.
	\end{aligned}
\end{equation}
Thus, each received codeword contributes a level-dependent log-likelihood
weight, and the analytical MAP rule is
\begin{equation}\label{eq:multibit_analytical_MAP_rule}
	\hat H_{\mathrm{MAP}}(\mathbf n)
	=
	\begin{cases}
		H_1, & \Lambda(\mathbf n)\geq 0,\\
		H_0, & \Lambda(\mathbf n)<0.
	\end{cases}
\end{equation}
Equation \eqref{eq:multibit_analytical_MAP_LLR} also makes explicit that the
fixed-dimensional statistic $\bar{\mathbf v}$ is sufficient when $K$ is known.
For unequal deployment priors, $K$ must be retained because it determines the
relative contribution of the accumulated evidence and the prior log-odds.
The rule in \eqref{eq:multibit_analytical_MAP_rule} is the analytical MAP
detector for the class-conditional categorical PMFs induced by a given
multi-bit quantizer. Their estimation under unknown observation statistics is
described next.
}

\textcolor{black}{
\section{Controller Training and Post-Training PMF Estimation}
\label{sec:training_pmf_estimation}
The proposed quantizer and detector are implemented in two distinct stages.
First, the neural probability controller is trained using the model-driven
objective derived for the corresponding quantization resolution. After the
controller is fixed, the class-conditional PMFs of its quantized outputs are
estimated from independent labeled data. \textcolor{black}{These estimated PMFs instantiate the analytical MAP rule for deployment.}
}

\subsection{\textcolor{black}{Probability Controller Training}}
\textcolor{black}{We consider a labeled training data set
$\mathcal D_{\mathrm{tr},i}=\{x_{t,i}^{\mathrm{tr}}\}_{t=1}^{T}$
for each $H_i$, where the conditional observation distributions are not
assumed to be analytically known. With mini-batch training
\cite{o2017introduction}, let $B$ denote the number of samples from each
hypothesis in a mini-batch. The two hypotheses are sampled separately so that
every update uses a class-balanced mini-batch. During training, we do not sample
the dithering variables or generate hard quantization outputs. Instead, the
controller outputs are used to compute the conditional output PMFs induced by
the probabilistic quantizer. This avoids the non-differentiability of hard
quantization while remaining consistent with the deployed quantizer, since
$G_{\Phi}(x)=P_{\Phi}(u=1\mid X=x)$ in the binary case and
$[\mathbf q_{\Phi}(x)]_l=P_{\Phi}(u=u_l\mid X=x)$ in the multi-bit case.}

\textcolor{black}{
For binary quantization, the class-conditional probability in
\eqref{eq:loss_func_quantizer} is approximated within each mini-batch as
\begin{equation}\label{eq:binary_training_PMF_estimate}
	\widehat{\gamma}_{\Phi,i}^{\mathrm{tr}}
	=
	\frac{1}{B}\sum_{t=1}^{B}G_{\Phi}(x_{t,i}^{\mathrm{tr}}),
	\quad i=0,1.
\end{equation}
Substituting \eqref{eq:binary_training_PMF_estimate} into
\eqref{eq:loss_func_quantizer} gives the empirical training objective
\begin{equation}\label{eq:binary_empirical_training_loss}
	\begin{aligned}
		\widehat{\mathcal L}_{\Phi}^{\mathrm{Binary}}
		=\sum_{k=0}^{K}C_K^k\min\{&
		\pi_0(\widehat{\gamma}_{\Phi,0}^{\mathrm{tr}})^k
		(1-\widehat{\gamma}_{\Phi,0}^{\mathrm{tr}})^{K-k},\\
		&
		\pi_1(\widehat{\gamma}_{\Phi,1}^{\mathrm{tr}})^k
		(1-\widehat{\gamma}_{\Phi,1}^{\mathrm{tr}})^{K-k}\}.
		\end{aligned}
\end{equation}
Thus, each training update first aggregates the controller outputs into the two
scalars $\widehat{\gamma}_{\Phi,0}^{\mathrm{tr}}$ and
$\widehat{\gamma}_{\Phi,1}^{\mathrm{tr}}$, and then evaluates the exact
finite-$K$ MAP detection error associated with these mini-batch estimates.
Since the empirical PMFs remain differentiable functions of $\Phi$, the
gradient of \eqref{eq:binary_empirical_training_loss} can be propagated directly
through $G_{\Phi}(\cdot)$.
}

\textcolor{black}{For multi-bit quantization, Proposition \ref{prop:tree_distribution} allows the
class-conditional PMF of the $L$ quantization levels to be approximated without
sampling the hard codewords:
\begin{equation}\label{eq:multibit_training_PMF_estimate}
	\widehat{\gamma}_{\Phi,i,l}^{\mathrm{tr}}
	=
	\frac{1}{B}\sum_{t=1}^{B}
	[\mathbf q_{\Phi}(x_{t,i}^{\mathrm{tr}})]_l,
	\quad
		i=0,1,\quad l=0,\ldots,L-1.
\end{equation}
Because every $\mathbf q_{\Phi}(x)$ is a valid PMF,
$\sum_{l=0}^{L-1}\widehat{\gamma}_{\Phi,i,l}^{\mathrm{tr}}=1$ for
$i=0,1$. Substituting these empirical PMFs into \eqref{eq:multibit_loss}
produces a differentiable surrogate objective. Since $\pi_0$, $\pi_1$, and $K$
are fixed during training, minimizing the bound in \eqref{eq:multibit_loss} is
equivalent to minimizing its log-domain Bhattacharyya coefficient
\begin{equation}\label{eq:multibit_empirical_training_loss}
	\widehat{\mathcal L}_{\Phi}^{\mathrm{Multi}}
	=
	\log\left(
	\sum_{l=0}^{L-1}
	\sqrt{
		\widehat{\gamma}_{\Phi,0,l}^{\mathrm{tr}}
		\widehat{\gamma}_{\Phi,1,l}^{\mathrm{tr}}
	}
	\right).
\end{equation}
Unlike the exact multi-bit MAPDEP in \eqref{eq:multi_bit_MAPDEP}, evaluating
\eqref{eq:multibit_empirical_training_loss} does not enumerate the
$\binom{K+L-1}{L-1}$ possible count vectors. Once the leaf probabilities have
been obtained, its evaluation requires only a summation over the $L$
quantization levels.}
\textcolor{black}{For either resolution, one training iteration therefore consists of evaluating the controller on the two class-balanced mini-batches, forming the empirical
class-conditional PMFs, computing the corresponding model-driven objective, and
updating $\Phi$ by backpropagation. An independent validation set is evaluated
periodically using the same resolution-specific objective, and the parameters
of the selected checkpoint are denoted by $\Phi^*$.}

\subsection{\textcolor{black}{Post-Training PMF Estimation and Plug-In MAP Fusion}}
\textcolor{black}{The analytical detectors in Sections \ref{subsec:binary_detector} and
\ref{subsec:multibit_detector} depend only on the class-conditional PMFs of the quantized
messages, rather than on the continuous observation densities. To estimate
these PMFs, we use an independent held-out data set
\begin{equation}
	\mathcal D_{\mathrm{pmf},i}
	=
	\{x_{t,i}^{\mathrm{pmf}}\}_{t=1}^{N_i},
	\quad i=0,1,
\end{equation}
after the controller $\Phi^*$ has been trained and fixed. This data set is not
used to update $\Phi^*$. We refer to this procedure as \emph{post-training PMF
estimation}. It estimates the distribution induced by the frozen quantizer and
does not calibrate or modify the controller outputs.}

\textcolor{black}{To present the binary and multi-bit cases in a unified form, define
$\mathbf r_{\Phi^*}(x)\in[0,1]^L$ as the distribution over the quantization
levels induced by the frozen controller, with
\begin{equation}\label{eq:unified_output_distribution}
	\mathbf r_{\Phi^*}(x)
	=
	\begin{cases}
		[1-G_{\Phi^*}(x),\,G_{\Phi^*}(x)]^T, & L=2,\\
		\mathbf q_{\Phi^*}(x), & L>2.
	\end{cases}
\end{equation}
The class-conditional probability of output level $u_l$ is estimated as
\begin{equation}\label{eq:unified_PMF_estimation}
	\widetilde{\gamma}_{i,l}
	=
	\frac{
		\displaystyle
		\sum_{t=1}^{N_i}
		[\mathbf r_{\Phi^*}(x_{t,i}^{\mathrm{pmf}})]_l+\alpha
	}{
		N_i+L\alpha
	},
	\quad
	i=0,1,\quad l=0,\ldots,L-1,
\end{equation}
where $\alpha>0$ is a symmetric smoothing parameter. Since every
$\mathbf r_{\Phi^*}(x)$ is a valid PMF,
$\sum_{l=0}^{L-1}\widetilde{\gamma}_{i,l}=1$, while the smoothing ensures
$\widetilde{\gamma}_{i,l}>0$ and prevents unbounded log-likelihood ratios.}

\textcolor{black}{Let $\mathbf n=[n_0,\ldots,n_{L-1}]^T$ denote the count vector of the received
quantization levels and define the estimated level-wise log-likelihood weights
as
\begin{equation}\label{eq:estimated_level_LLR_weight}
	\widetilde w_l
	\triangleq
	\log\frac{\widetilde{\gamma}_{1,l}}
	          {\widetilde{\gamma}_{0,l}},
	\qquad
	\widetilde{\mathbf w}
	=[\widetilde w_0,\ldots,\widetilde w_{L-1}]^T.
\end{equation}
Substituting the estimated PMFs into the analytical MAP detector yields the
plug-in log-posterior odds
\begin{equation}\label{eq:unified_plugin_MAP_LLR}
	\widehat{\Lambda}(\mathbf n)
	=
	\log\frac{\pi_1}{\pi_0}
	+\sum_{l=0}^{L-1}n_l\widetilde w_l
	=
	\log\frac{\pi_1}{\pi_0}
	+\mathbf n^T\widetilde{\mathbf w}.
\end{equation}
The plug-in MAP decision is
\begin{equation}\label{eq:unified_plugin_MAP_rule}
	\widehat H_{\mathrm{plug\text{-}in}}
	=
	\begin{cases}
		H_1, & \widehat{\Lambda}(\mathbf n)\geq 0,\\
		H_0, & \widehat{\Lambda}(\mathbf n)<0.
	\end{cases}
\end{equation}
For the binary case, $\mathbf n=[K-s,s]^T$, and
\eqref{eq:unified_plugin_MAP_LLR} reduces to
\eqref{eq:binary_analytical_MAP_LLR} after replacing the true Bernoulli
probabilities by their estimates. For the multi-bit case,
$\mathbf n=K\bar{\mathbf v}$, and \eqref{eq:unified_plugin_MAP_LLR} is the
estimated counterpart of \eqref{eq:multibit_analytical_MAP_LLR}. If the
estimated PMFs equal the true class-conditional PMFs, the plug-in rule
coincides with the corresponding analytical MAP detector. Otherwise, it
provides a data-driven implementation of that detector under unknown
observation statistics. The deployment priors $\pi_0$ and $\pi_1$ are
specified independently of the class proportions in the training and PMF
estimation data sets.}

\section{Simulation Results}\label{sec:simulation}
\begin{figure*}[!t]
    \centering
    \subfloat[$M=1$ bit\label{subfig:true_pdf_m1}]{
        \includegraphics[width=0.4\textwidth]{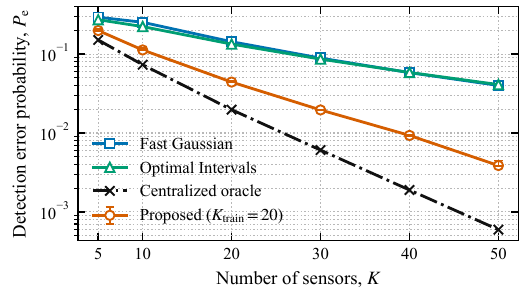}
    }\hspace{0.03\textwidth}
    \subfloat[$M=2$ bits\label{subfig:true_pdf_m2}]{
        \includegraphics[width=0.4\textwidth]{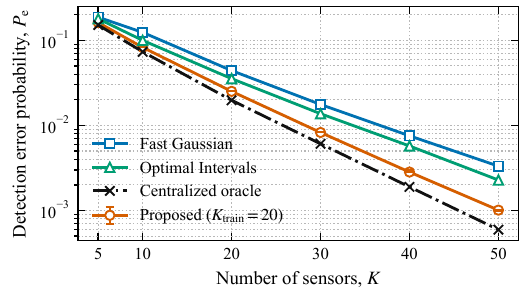}
    }

    \vspace{-0.35em}

    \subfloat[$M=3$ bits\label{subfig:true_pdf_m3}]{
        \includegraphics[width=0.4\textwidth]{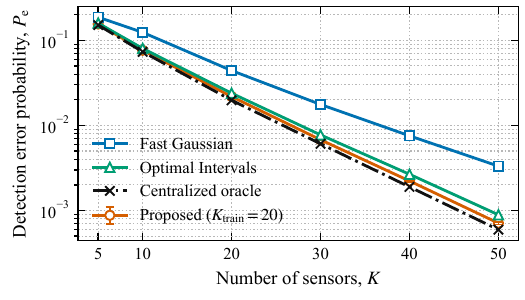}
    }\hspace{0.03\textwidth}
    \subfloat[$M=4$ bits\label{subfig:true_pdf_m4}]{
        \includegraphics[width=0.4\textwidth]{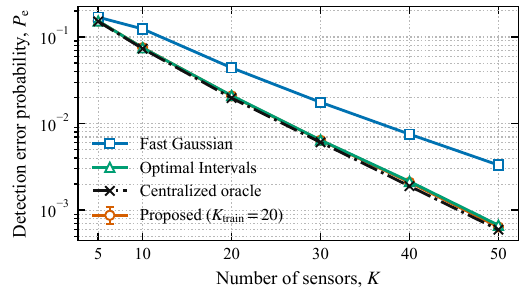}
    }
    \caption{\textcolor{black}{Detection error probability versus the number of
    sensors when the classical baselines know the true random-polarity PDFs. The proposed
    controller is trained at $K_{\mathrm{train}}=20$ and frozen for all
    deployment values of $K$.}}
    \label{fig:k_sweep_true_pdf}
\end{figure*}

\textcolor{black}{Motivated by the detection of undecodable zero-mean modulated signals in spectrum sensing \cite{sahai2004fundamental,tandra2008snr}, we consider a distributed random-polarity signal-presence problem. The local observation
at sensor $k$ is modeled as
\begin{equation}
    \begin{aligned}
        H_0 &: X_k=N_k,\\
        H_1 &: X_k=S_kA+N_k,
    \end{aligned}
\end{equation}
where $A>0$ is the signal amplitude, $S_k\in\{-1,+1\}$ is equiprobable,
and $N_k\sim\mathcal N(0,1)$. $S_k$ is independent of $N_k$. The pairs
$\{S_k,N_k\}_{k=1}^{K}$ are independent across sensors. Equivalently,
\begin{align}
    f_0(x)
    &=\mathcal N(x;0,1),\\
    f_1(x)
    &=\frac{1}{2}\mathcal N(x;-A,1)
      +\frac{1}{2}\mathcal N(x;A,1).
\end{align}}

\textcolor{black}{This model is a scalar abstraction of detecting the presence of an antipodal or BPSK-type transmission when its instantaneous data symbols are not decoded. Under $H_0$, each sensor observes noise only, whereas under $H_1$, its matched-filter output contains an unknown-polarity signal component. The independent variables $S_k$ represent independently varying local symbol realizations and retain the conditionally i.i.d. observation model considered in our analysis.}

\textcolor{black}{The probability controllers are trained with $K_{\mathrm{train}}=20$ sensors. \textcolor{black}{Each controller is a fully connected network with three 20-unit ReLU hidden layers and a sigmoid output layer of dimension one for $M=1$ and $2^M-1$ for $M\geq2$.} For each hypothesis, \num{50000} labeled local observations are generated and partitioned into \num{30000} training observations, \num{10000} validation observations, and \num{10000} observations for PMF estimation. Each controller is trained for 500 epochs using the Adam optimizer \cite{kingma2014adam} with a constant learning rate of $10^{-4}$. Each mini-batch contains \num{1000} local observations from each hypothesis, for a total batch size of \num{2000}. For the observation model, unless otherwise stated, we use equal priors $\pi_0=\pi_1=0.5$ and a signal amplitude of $A=1.4111$, which corresponds to approximately $3$~dB local SNR.}

\textcolor{black}{After training, the controller is fixed, and the class-conditional PMFs of its quantized outputs are estimated from the held-out PMF estimation set using a smoothing parameter of $\alpha=1$. \textcolor{black}{The resulting level-wise log-likelihood ratios implement the plug-in MAP detector at the FC.} Each test configuration uses at least \num{100000} independent detection instances under each hypothesis; the amplitude-mismatch experiment uses \num{200000} per hypothesis to improve resolution at its low-error operating points. \textcolor{black}{Within each test configuration, all methods use the same generated observations. For methods involving training or data fitting, we perform five independent runs. Unless otherwise stated, the reported values are medians across these runs, and error bars indicate the interquartile range.} The proposed controllers are implemented and trained in PyTorch \cite{paszke2019pytorch}.}

\subsection{\textcolor{black}{Detection Performance across Network Sizes}}

\begin{figure*}[!t]
    \centering
    \subfloat[$M=1$ bit\label{subfig:sample_fitted_m1}]{
        \includegraphics[width=0.4\textwidth]{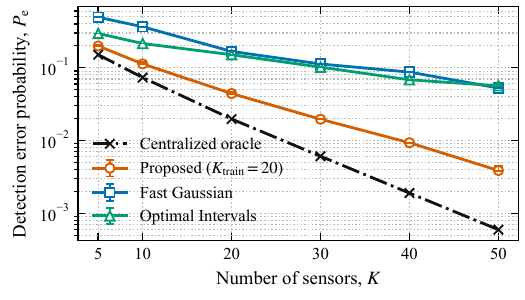}
    }\hspace{0.03\textwidth}
    \subfloat[$M=2$ bits\label{subfig:sample_fitted_m2}]{
        \includegraphics[width=0.4\textwidth]{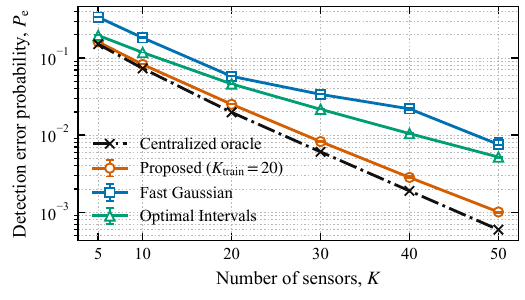}
    }

    \vspace{-0.35em}

    \subfloat[$M=3$ bits\label{subfig:sample_fitted_m3}]{
        \includegraphics[width=0.4\textwidth]{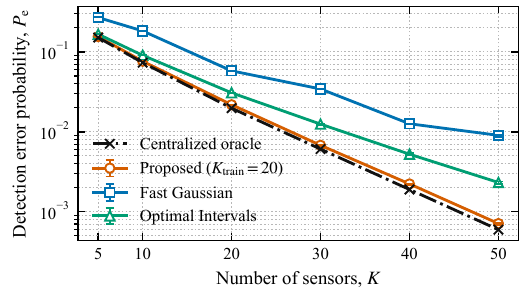}
    }\hspace{0.03\textwidth}
    \subfloat[$M=4$ bits\label{subfig:sample_fitted_m4}]{
        \includegraphics[width=0.4\textwidth]{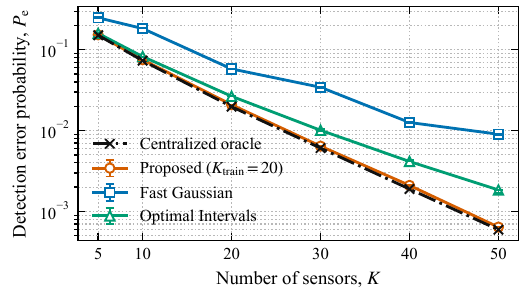}
    }
    \caption{\textcolor{black}{Detection error probability versus the number of
    sensors when the observation distribution is unknown. The classical baselines fit
    one Gaussian per hypothesis using the corresponding \num{30000} training
    observations; all other settings are unchanged.}}
    \label{fig:k_sweep_sample_fitted}
\end{figure*}

\textcolor{black}{\textcolor{black}{Figs.~\ref{fig:k_sweep_true_pdf} and
\ref{fig:k_sweep_sample_fitted} evaluate the detection performance as the
deployment network size $K$ varies, for message resolutions $M=1,\ldots,4$.
The proposed method is compared with Fast Gaussian \cite{gul2021fast},
Optimal Intervals \cite{altay2016optimal}, and the unquantized centralized
oracle.} Fast
Gaussian provides a computationally efficient classical design based on a
Gaussian approximation of the aggregate quantized statistic. Optimal
Intervals optimizes an ordered interval quantizer using Chernoff
information and applies its native fusion rule. The centralized oracle has
access to the raw observations and therefore serves as an unquantized
performance reference rather than a communication-constrained competitor.
\textcolor{black}{It uses the exact continuous-observation likelihood in both comparison settings.}
Fast Gaussian is designed for each deployment value of $K$, whereas the
Optimal Intervals quantizer is $K$-invariant.}

\textcolor{black}{Fig.~\ref{fig:k_sweep_true_pdf} presents the more
informative baseline condition in which both classical quantizer designs know
the exact random-polarity PDFs. The error probability decreases with $K$ for every method,
showing the accumulation of distributed sensing information at the FC. The
proposed method is particularly effective in the low-rate regime. At
$K=20$, its error probability is $0.0442$ with one-bit messages,
compared with $0.1435$ for Fast Gaussian and $0.1339$ for Optimal
Intervals. With two-bit messages, the corresponding values are $0.0251$,
$0.0441$, and $0.0354$, respectively. This advantage is consistent with the
two-sided random-polarity likelihood ratio: observations in both tails support
$H_1$, so an effective low-rate controller must reuse a message
label over disconnected observation regions.}

\textcolor{black}{As the message resolution increases, the proposed
controller retains progressively more of the local likelihood information
and approaches the unquantized oracle. For example, at $K=20$ and $M=4$, the
proposed, Optimal Intervals, and centralized-oracle error probabilities
are $0.0209$, $0.0210$, and $0.0197$, respectively. At $K=50$, the
corresponding values are $6.4\times10^{-4}$,
$6.7\times10^{-4}$, and $5.95\times10^{-4}$. Thus, the largest improvement
is obtained when moving from one to two bits, while the gap between the
proposed quantized detector and the centralized oracle becomes small at
three and four bits.}

\textcolor{black}{In this experiment, we also examine the finite-$K$ tightness of the training surrogate discussed in Remark~\ref{rem:multibit_surrogate_scope} by comparing the exact MAP error of the learned quantizers with the optimized Chernoff and Bhattacharyya bounds for $M=2,3$ and $K=5,10,20$. 
For each quantizer, all three quantities are computed from the same output PMFs obtained by numerical integration of the class-conditional observation densities.
Across these settings, the ratio of the optimized Chernoff bound to the exact MAP error ranges from $1.915$ to $3.013$.
$\lambda^\star$ is $0.536$ for $M=2$ and $0.540$ for $M=3$; optimizing
$\lambda$ reduces the Bhattacharyya bound by only approximately
$0.2\%$--$1.3\%$. Thus, the main finite-$K$ gap in these settings comes
from the upper bound itself rather than the fixed choice $\lambda=1/2$.
This numerically supports the interpretation in Remark~\ref{rem:multibit_surrogate_scope} of the training
objective as a conservative upper-bound surrogate for the finite-sensor
MAP error.}

\textcolor{black}{Fig.~\ref{fig:k_sweep_sample_fitted} considers the
practical case in which the classical baselines do not know the observation
law. \textcolor{black}{Because these baselines require specified PDFs, we provide them with one class-conditional Gaussian fitted from the training observations under each hypothesis.} This fitted model does not reproduce the
random-polarity mixture likelihood over the full observation range, and the
resulting mismatch degrades both classical designs. At $K=20$, their error
probabilities are $0.1675$ and $0.1509$ for $M=1$, compared with $0.0442$
for the proposed method. For $M=4$, the proposed method achieves $0.0209$,
whereas Fast Gaussian and Optimal Intervals achieve $0.0579$ and
$0.0266$, respectively. The same ordering persists as $K$ increases; at
$K=50$ and $M=4$, the respective error probabilities are
$6.4\times10^{-4}$, $9.03\times10^{-3}$, and
$1.85\times10^{-3}$. These results show that the learned controller can
exploit the non-Gaussian, two-sided structure directly from labeled local
observations, without access to an analytic PDF. We further verify it by examining the learned controller structure and the discrete LLRs used by the analytical fusion rule in Section~\ref{subsec:controller_structure}.}

\begin{figure*}[!t]
    \centering
    \subfloat[$M=1$ bit\label{subfig:controller_prob_m1}]{
        \includegraphics[width=0.4\textwidth]{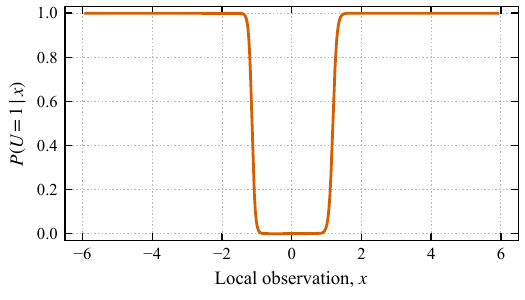}
    }\hspace{0.03\textwidth}
    \subfloat[$M=2$ bits\label{subfig:controller_prob_m2}]{
        \includegraphics[width=0.4\textwidth]{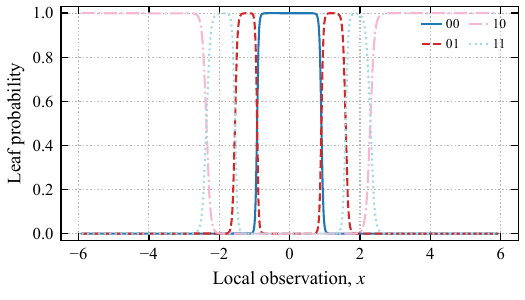}
    }

    \vspace{-0.35em}

    \subfloat[$M=3$ bits\label{subfig:controller_prob_m3}]{
        \includegraphics[width=0.4\textwidth]{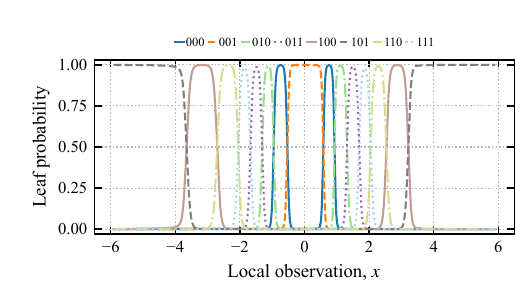}
    }\hspace{0.03\textwidth}
    \subfloat[$M=4$ bits\label{subfig:controller_prob_m4}]{
        \includegraphics[width=0.4\textwidth]{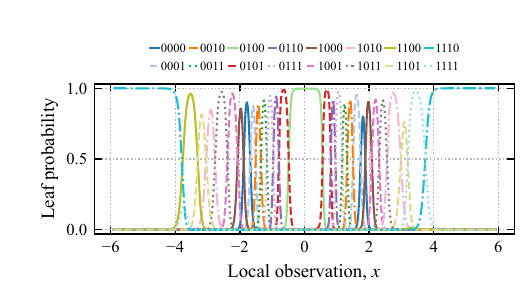}
    }
    \caption{\textcolor{black}{Representative learned controller maps. The curves show the binary output probability for $M=1$ and
    the tree-leaf codeword probabilities for $M=2,3,4$.}}
    \label{fig:controller_probabilities}
\end{figure*}

\begin{figure*}[!t]
    \centering
    \subfloat[$M=1$ bit\label{subfig:controller_llr_m1}]{
        \includegraphics[width=0.4\textwidth]{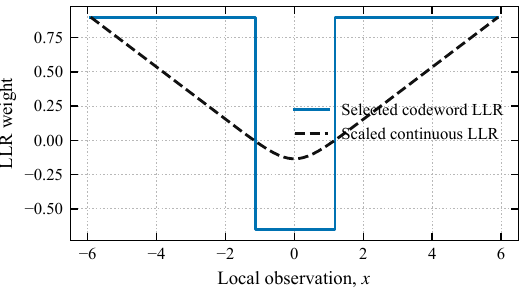}
    }\hspace{0.03\textwidth}
    \subfloat[$M=2$ bits\label{subfig:controller_llr_m2}]{
        \includegraphics[width=0.4\textwidth]{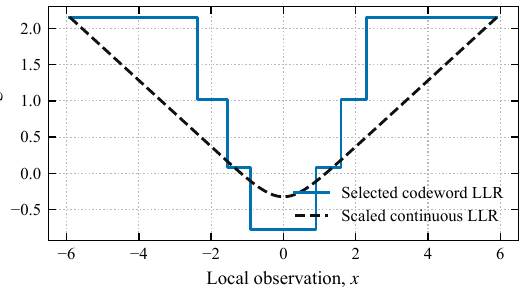}
    }

    \vspace{-0.35em}

    \subfloat[$M=3$ bits\label{subfig:controller_llr_m3}]{
        \includegraphics[width=0.4\textwidth]{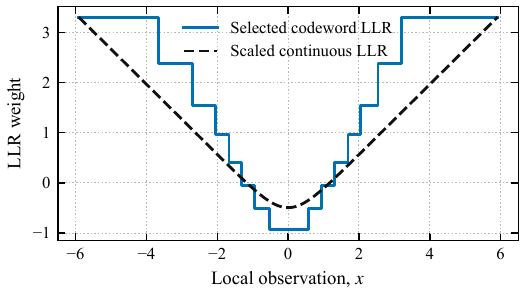}
    }\hspace{0.03\textwidth}
    \subfloat[$M=4$ bits\label{subfig:controller_llr_m4}]{
        \includegraphics[width=0.4\textwidth]{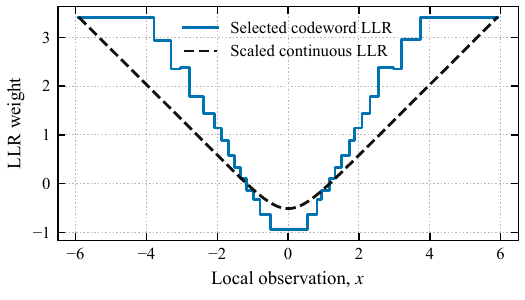}
    }
    \caption{\textcolor{black}{Discrete LLR representation induced by the
    learned controllers in Fig.~\ref{fig:controller_probabilities}. At
    each observation value, the solid curve is the estimated LLR weight of
    the maximum-probability codeword, and the dashed curve is the continuous random-polarity
    LLR scaled to the same range.}}
    \label{fig:controller_llr}
\end{figure*}

\begin{figure}[!t]
    \centering
    \subfloat[Fully frozen\label{subfig:amplitude_frozen}]{
        \includegraphics[width=0.92\columnwidth]{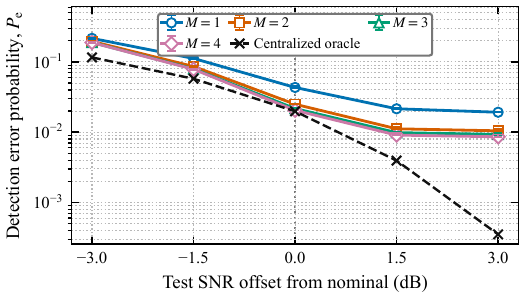}
    }

    \vspace{-0.45em}

    \subfloat[FC recalibration\label{subfig:amplitude_recalibrated}]{
        \includegraphics[width=0.92\columnwidth]{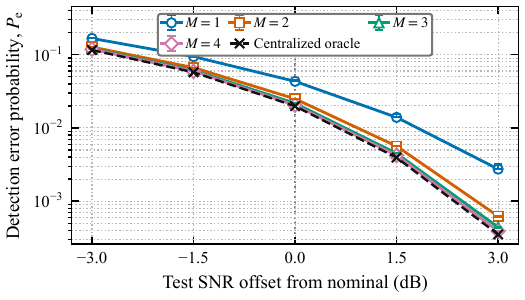}
    }
    \caption{\textcolor{black}{Robustness to a test-time signal-amplitude
    mismatch at $K=20$. The local controller remains frozen at the nominal
    amplitude. Panels (a) and (b) respectively use the nominal FC output
    PMFs and PMFs re-estimated from \num{10000} target-amplitude observations
    per hypothesis. The centralized oracle is independent of $M$.}}
    \label{fig:amplitude_mismatch}
\end{figure}

\begin{figure}[!t]
    \centering
    \subfloat[Frozen i.i.d. FC\label{subfig:heterogeneity_frozen}]{
        \includegraphics[width=0.92\columnwidth]{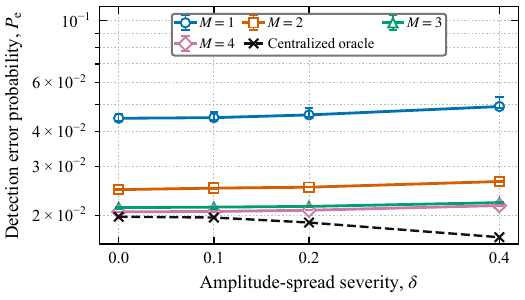}
    }

    \vspace{-0.45em}

    \subfloat[Sensor-specific FC\label{subfig:heterogeneity_specific}]{
        \includegraphics[width=0.92\columnwidth]{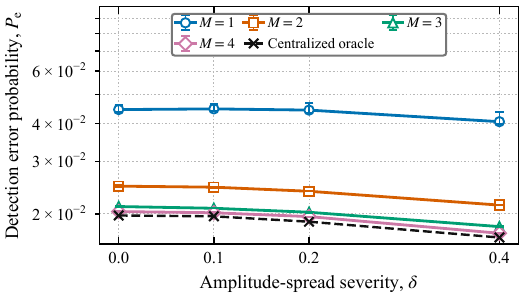}
    }
    \caption{\textcolor{black}{Robustness to heterogeneous sensor amplitudes
    at $K=20$. The amplitude-spread parameter $\delta$ ranges from $0$ to
    $0.4$ while the root-mean-square amplitude is held fixed. Panels (a) and
    (b) respectively use the frozen i.i.d. FC and sensor-specific output
    likelihoods without retraining the common controller.}}
    \label{fig:heterogeneity}
\end{figure}

\begin{figure}[!t]
    \centering
    \subfloat[Equal-prior FC\label{subfig:prior_equal}]{
        \includegraphics[width=0.92\columnwidth]{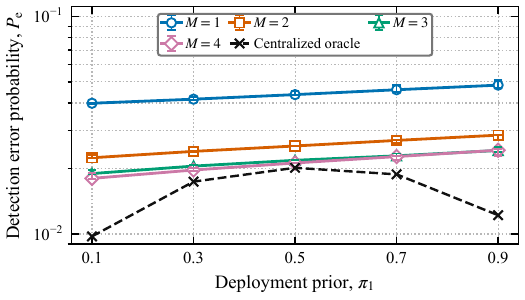}
    }

    \vspace{-0.45em}

    \subfloat[Correct-prior FC\label{subfig:prior_correct}]{
        \includegraphics[width=0.92\columnwidth]{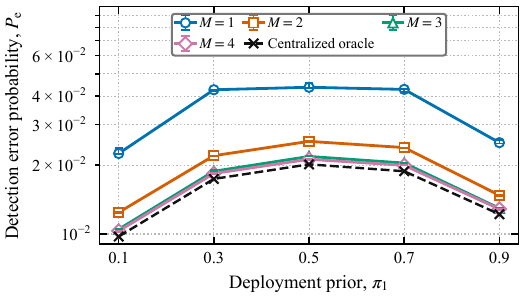}
    }
    \caption{\textcolor{black}{Generalization to unequal deployment priors at
    $K=20$. Panels (a) and (b) respectively retain the equal-prior FC and
    update the hypothesis-prior term in the MAP rule. The local controller
    and class-conditional output PMFs remain fixed.}}
    \label{fig:unequal_prior}
\end{figure}

\begin{figure}[!t]
    \centering
    \subfloat[$K=20$\label{subfig:ablation_k20}]{
        \includegraphics[width=0.92\columnwidth]{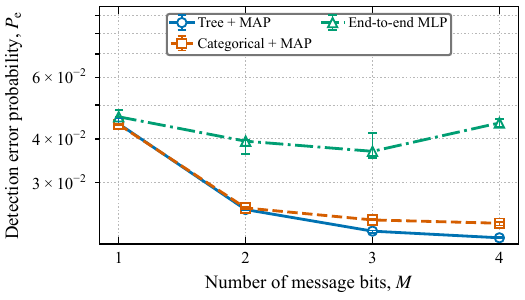}
    }

    \vspace{-0.45em}

    \subfloat[$K=50$\label{subfig:ablation_k50}]{
        \includegraphics[width=0.92\columnwidth]{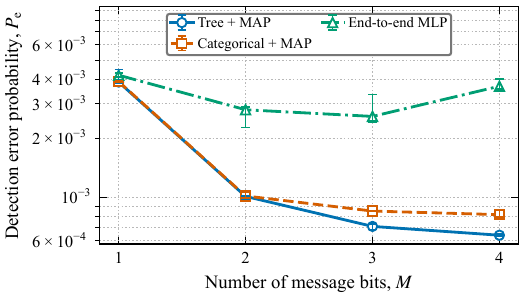}
    }
    \caption{\textcolor{black}{Three-method ablation across message resolutions
    at (a) $K=20$ and (b) $K=50$. Tree~+~MAP denotes the proposed pipeline:
    it uses the binary controller at $M=1$ and the tree-structured controller
    for $M\geq2$.
    Categorical~+~MAP \textcolor{black}{replaces the proposed controller} by a direct categorical
    controller while retaining the same model-driven objective and MAP fusion,
    whereas End-to-end MLP replaces the complete model-driven pipeline by
    joint data-driven training.}}
    \label{fig:stepwise_ablation}
\end{figure}

\textcolor{black}{These two settings also illustrate the trade-off between model knowledge and computation. When accurate PDFs are available, the classical interval quantizers avoid neural training and require only threshold comparisons at each sensor, whereas the proposed method incurs offline controller training and a DNN forward pass followed by an $M$-step tree traversal. At the FC, all considered quantized methods remain lightweight, requiring approximately $\mathcal O(K)$ likelihood accumulation, and they have the same $M$-bit communication cost per sensor. Thus, the proposed method exchanges additional offline and sensor-side computation for the ability to learn from samples without an explicitly specified observation model. }

\subsection{\textcolor{black}{Learned Controller Structure}}
\label{subsec:controller_structure}

\textcolor{black}{Fig.~\ref{fig:controller_probabilities} visualizes the learned quantization mechanism. 
For $M=1$, the controller assigns the same message label to the two tails and the other label to the central region. 
It therefore realizes a non-contiguous decision region that cannot be represented by a single monotone threshold in $x$. 
For multiple bits, the tree leaves form progressively finer, approximately symmetric soft partitions of the observation axis. 
Most leaf probabilities are close to zero or one away from transition regions, while the smooth transitions retain the probabilistic controller used during training.}

\textcolor{black}{Fig.~\ref{fig:controller_llr} shows how these controller
maps translate into the statistics used at the FC. The selected codeword
has a negative LLR weight near the origin and increasingly positive weights
in both tails, matching the even, two-sided shape of the continuous
random-polarity LLR.
The one-bit controller gives a two-level approximation, whereas increasing
$M$ produces a finer staircase that follows the continuous LLR more
closely. Hence, the learned local messages preserve the detection-relevant
ordering in $|x|$, and the estimated level-wise weights provide the
corresponding plug-in MAP fusion rule.}

\subsection{\textcolor{black}{Robustness Under Non-Ideal Deployment}}

\textcolor{black}{Fig.~\ref{fig:amplitude_mismatch} varies the deployment
SNR by $\pm3$ dB around the nominal training condition. Keeping both the
controller and the nominal FC weights frozen is operational, but becomes
increasingly mismatched away from the nominal point. FC recalibration re-estimates only the level-wise PMFs. This step substantially recovers the lost performance
without retraining or changing the sensor controller. At $M=4$, for
example, recalibration reduces $P_e$ from $0.1889$ to $0.1189$ at an SNR offset of $-3$ dB
and from $8.60\times10^{-3}$ to $3.90\times10^{-4}$ at an offset of $+3$ dB; the
corresponding unquantized-oracle values are $0.1162$ and
$3.40\times10^{-4}$. The improvement from FC-only recalibration is already
pronounced at $M=2$ and becomes nearly oracle-level at three and four bits.}

\textcolor{black}{Fig.~\ref{fig:heterogeneity} evaluates a departure from
the identically distributed sensor model. The sensor amplitudes are spread
linearly around the nominal value and normalized to preserve their
root-mean-square value. \textcolor{black}{Specifically, $A_k=A c_k/\sqrt{K^{-1}\sum_{j=1}^{K}c_j^2}$, where $c_k=1+\delta\left(2(k-1)/(K-1)-1\right)$. The sensor-specific FC uses the MAP score $\log(\pi_1/\pi_0)+\sum_{k=1}^{K}\log[p_k(u_k\mid H_1)/p_k(u_k\mid H_0)]$, where $p_k(u_k\mid H_i)$ is the output PMF induced by the frozen shared controller at amplitude $A_k$.} \textcolor{black}{In this experiment, these PMFs are computed by numerically integrating the frozen controller's conditional output probabilities against the known sensor-specific observation densities, using the amplitudes $A_k$, unit-variance Gaussian noise, and equiprobable random-polarity model. The FC is provided with these sensor-specific PMFs and the sensor index associated with each received message.} Under the frozen i.i.d. FC, increasing heterogeneity
causes a modest loss; at $\delta=0.4$, $P_e$ is $0.0492$, $0.0265$,
$0.0222$, and $0.0217$ for $M=1,\ldots,4$. Retaining the same shared
controller but using sensor-specific likelihood weights at the FC reduces
these values to $0.0407$, $0.0214$, $0.0181$, and $0.0172$,
respectively. Thus, deployment heterogeneity can be handled primarily at
the analytical fusion stage, and the remaining gap to the unquantized
oracle is small for three- and four-bit messages.}

\textcolor{black}{Fig.~\ref{fig:unequal_prior} varies the deployment prior
$\pi_1=P(H_1)$ from $0.1$ to $0.9$. Using the correct prior in the
analytical MAP threshold consistently improves over retaining the equal-prior
rule away from $\pi_1=0.5$. At $\pi_1=0.1$, for example, the weighted error
at $M=4$ decreases from $0.0180$ to $0.0103$, close to the unquantized
oracle value $0.00979$; at $\pi_1=0.9$, it decreases from $0.0243$ to
$0.0129$, compared with $0.0122$ for the oracle. This adaptation requires
neither new local observations nor controller retraining, because the
class-conditional output PMFs are unchanged and only the prior log-odds in
the MAP rule must be updated.}

\subsection{\textcolor{black}{Ablation Study}}

\textcolor{black}{We compare three pipelines. Tree~+~MAP is the proposed controller with plug-in MAP fusion. Categorical~+~MAP uses the same three-layer, 20-unit backbone but replaces the structured controller output with an $L$-way softmax; it retains the same resolution-specific model-driven objective, PMF estimation, and MAP fusion. End-to-end MLP jointly trains this direct categorical controller and a fusion MLP with three 30-unit ReLU hidden layers using cross-entropy and straight-through Gumbel--Softmax messages; the fusion input is the message average for $M=1$ and the $L$-bin message histogram for $M\geq2$.}

\textcolor{black}{Fig.~\ref{fig:stepwise_ablation} evaluates the proposed design
at two levels. For $M\geq2$, the Tree~+~MAP versus Categorical~+~MAP comparison
isolates the tree-structured controller because both methods use the same
model-driven training objective and analytical MAP fusion. The $M=1$ point
uses the binary controller rather than a tree and is included only for the
comparison of the complete model-driven pipeline with the End-to-end MLP.
\textcolor{black}{All methods use the same total labeled-data budget and test observations.}
At $K=20$, the proposed pipeline achieves
error probabilities $0.0442$, $0.0251$, $0.0218$, and $0.0209$ for
$M=1,\ldots,4$. \textcolor{black}{Under the tested random-polarity observation model and training configuration,} direct categorical training is close
at $M=2$ but is worse at three and four bits, where its error probabilities are $0.0235$ and
$0.0230$. The End-to-end MLP provides a purely data-driven comparison with the
complete proposed pipeline and does not benefit reliably from additional
message bits; at $K=20$ and $M=4$, its error probability is $0.0445$. The
differences become more pronounced as $K$ grows: at $K=50$ and $M=4$, the
error probabilities are $6.4\times10^{-4}$ for Tree~+~MAP,
$8.15\times10^{-4}$ for Categorical~+~MAP, and
$3.69\times10^{-3}$ for End-to-end MLP. Because the end-to-end baseline changes
both the controller-training objective and the fusion rule, this comparison
supports the complete MAP-guided model-driven design rather than attributing
the performance gap to either component alone.}

\section{Conclusion}\label{sec:conclusion}
In this paper, we proposed a comprehensive model-driven DL framework for distributed detection over communication-constrained networks, encompassing both binary and multi-bit quantization schemes. We first established the theoretical foundations by \textcolor{black}{deriving the exact finite-sensor MAP detection error for fixed local quantizers}. A key theoretical finding is that, \textcolor{black}{\textcolor{black}{under conditionally i.i.d. observations, independent local randomizations, and a common quantization-level constraint, identical quantizers preserve the supremum of the average network Chernoff information}}. Guided by these insights, we designed a probabilistic quantization architecture. For the binary case, we directly optimized the exact MAP detection error probability. To address the computational intractability in higher-resolution scenarios, we extended the design to a scalable tree-structured multi-bit quantizer, employing a computationally efficient Bhattacharyya surrogate objective. Furthermore, we proved that \textcolor{black}{under the adopted identical-quantizer architecture, the binary message average and the multi-bit message histogram preserve the minimum MAP detection error} and derived analytical MAP fusion rules for both binary and multi-bit messages. \textcolor{black}{Under unknown observation statistics, the required class-conditional PMFs are estimated from held-out data after controller training and substituted into the plug-in MAP rule.} \textcolor{black}{Simulation results demonstrated that the proposed method outperforms the considered classical quantizer designs on the non-Gaussian random-polarity task and can accommodate amplitude mismatch, sensor heterogeneity, and unequal deployment priors through lightweight adaptation of the analytical fusion rule. \textcolor{black}{In the tested settings, the ablation results show a benefit of the tree-structured controller over the direct categorical controller at three and four bits, and support the complete MAP-guided model-driven design in comparison with the considered end-to-end baseline.}}\vspace{-1em}

\FloatBarrier
\appendices
\section{Proof of Lemma \ref{lem:glob_opt_identical_quantizer}}\label{append:proof_identical_quantizer}
\textcolor{black}{Consider an arbitrary collection of possibly heterogeneous local quantizers $\{p(u_k|X_k)\}_{k=1}^K$. Let 
\begin{equation}
	p_k(u|H_i)\triangleq \mathbb{E}_{X|H_i}[p(u_k=u\mid X)],\ \ i\in\{0,1\},
\end{equation}
denote the conditional probability mass function of the quantized message generated by sensor $k$. For any $\lambda\in[0,1]$, define 
\begin{equation}
	Z_k(\lambda)\triangleq\sum_{u=0}^{L-1}p_k(u|H_0)^{\lambda}p_k(u|H_1)^{1-\lambda}.
\end{equation}
Since the local observations are conditionally independent and the local quantization randomizations are mutually independent \textcolor{black}{and independent of $(H,X_1,\ldots,X_K)$}, the quantized messages are conditionally independent given $H_i$. Therefore,
\begin{equation}
	p(\mathbf{u}|H_i)=\prod_{k=1}^{K}p_k(u_k|H_i),\ i=0,1.
\end{equation} 
It follows that 
\begin{equation}
	\begin{aligned}
		&\sum_{\mathbf{u}}p(\mathbf{u}|H_0)^{\lambda}p(\mathbf{u}|H_1)^{1-\lambda}\\
		&=\prod_{k=1}^{K}\left[\sum_{u_k=0}^{L-1}p_k(u_k|H_0)^{\lambda}p_k(u_k|H_1)^{1-\lambda}\right]\\
		&=\prod_{k=1}^{K}Z_k(\lambda).
	\end{aligned}
\end{equation}
Hence, the average network Chernoff information is 
\begin{equation}
	C(H,\mathbf{u})=-\frac{1}{K}\min_{0\leq\lambda\leq 1}\sum_{k=1}^{K}\log Z_k(\lambda).
\end{equation}
Using 
\begin{equation}
	\min_{\lambda}\sum_{k=1}^{K}f_k(\lambda)\geq\sum_{k=1}^{K}\min_{\lambda}f_k(\lambda),
\end{equation}
we obtain 
\begin{equation}
	C(H,\mathbf{u})\leq-\frac{1}{K}\sum_{k=1}^{K}\min_{0\leq\lambda\leq1}\log Z_k(\lambda)=\frac{1}{K}\sum_{k=1}^{K}C_k,
\end{equation}
where 
\begin{equation}
	C_k\triangleq -\min_{0\leq\lambda\leq 1}\log Z_k(\lambda)
\end{equation}
is the single-sensor Chernoff information induced by the quantizer $p(u_k|X_k)$. Define the \textcolor{black}{supremum of the} single-sensor Chernoff information as
\begin{equation}
	C^*\triangleq\textcolor{black}{\sup_{p(u|X)}}\left\{-\min_{0\leq\lambda\leq 1}\log\left[\sum_{u=0}^{L-1}p(u|H_{0})^{\lambda}p(u|H_{1})^{1-\lambda}\right]\right\}.
\end{equation}
Since every sensor is subject to the same $L$-level quantization constraint, $C_k\leq C^*$ for all $k$. Therefore, every possibly heterogeneous quantizer collection satisfies
\begin{equation}\label{eq:global_Chernoff_bound}
	C(H,\mathbf{u})\leq\frac{1}{K}\sum_{k=1}^{K}C_k\leq C^*.
\end{equation}
\textcolor{black}{Conversely, let $p(u|X)$ be any admissible single-sensor quantizer, and deploy it at every sensor with independent local randomizations:
\begin{equation}
	p(u_k|X_k)=p(u|X),\ \ k=1,\ldots,K.
\end{equation}
The induced conditional message distributions are then identical across sensors. Denoting their common Chernoff coefficient by
\begin{equation}
	Z(\lambda)=\sum_{u=0}^{L-1}p(u|H_0)^{\lambda}p(u|H_1)^{1-\lambda},
\end{equation}
we have
\begin{equation}
	\begin{aligned}
		C(H,\mathbf{u})=&-\frac{1}{K}\min_{0\leq\lambda\leq1}K\log Z(\lambda)\\
		=&-\min_{0\leq\lambda\leq 1}\log Z(\lambda).
	\end{aligned}
\end{equation}
Taking the supremum over the shared quantizer $p(u|X)$ gives $C^*$. Together with the upper bound in \eqref{eq:global_Chernoff_bound}, this establishes \eqref{eq:max_Chernoff}. If a single-sensor quantizer $p^*(u|X)$ attains $C^*$, its identical deployment also attains this global upper bound.} This completes the proof.}

\section{Proof of Proposition \ref{prop:MAPDEP_binary_quantize}}\vspace{-0.5em}
\label{append:proof_MAPDEP_binary_quantize}
Define $\mathcal{U}=\{\mathbf{u}=[u_1,\ldots,u_K]^T\mid u_k\in\{0,1\},\ k=1,\ldots,K\}=\{0,1\}^K$ and the sequence of sets $\{\mathcal{U}_k\}_{k=0}^K$, where $\mathcal{U}_k=\{\mathbf{u}\in\mathcal{U}\mid\frac{1}{K}\sum_{i=1}^K u_i=k/K\}$. If the binary quantized data $u_1,u_2,\ldots,u_K$ are conditionally i.i.d. given $H$, then every $\mathbf{u}\in\mathcal{U}_k$ has the same conditional probability given $H$, i.e.,
\begin{equation}\label{eq:bernoulli_distribution}
	\begin{aligned}		 
		p(\mathbf{u}|H)
		=&(p(u=1|H))^k(p(u=0|H))^{K-k}\\
		=&(\gamma(H))^k(1-\gamma(H))^{K-k},
	\end{aligned}
\end{equation}   
$\forall \mathbf{u}\in \mathcal{U}_k$ and $H\in\{H_0,H_1\}$, where $\gamma(H)\triangleq p(u=1|H)$ denotes the conditional probability that a quantized message equals `1' given $H$. Using \eqref{eq:bernoulli_distribution} and $|\mathcal{U}_k|=C_K^k$, the MAPDEP of binary-quantization-based detection is
\begin{equation}
	\label{eq:binary_MAPDEP_proof}
	\begin{aligned}
		P_{E}(H,\mathbf{u})
		=&\sum_{\mathbf{u}}\min\left\{\pi_0p(\mathbf{u}|H_0),\pi_1p(\mathbf{u}|H_1)\right\}\\
		=&\sum_{k=0}^{K}\sum_{\mathbf{u}\in\mathcal{U}_k}\min\left\{\pi_0p(\mathbf{u}|H_0),\pi_1p(\mathbf{u}|H_1)\right\}\\
		=&\sum_{k=0}^{K}C_{K}^{k}\min\{\pi_0(\gamma(H_0))^k(1-\gamma(H_0))^{K-k},\\
		&\pi_1(\gamma(H_1))^k(1-\gamma(H_1))^{K-k}\}.
	\end{aligned}
\end{equation} 
This completes the proof of Proposition \ref{prop:MAPDEP_binary_quantize}.\vspace{-1em}

\section{Proof of Theorem \ref{th:opt_mean_fusion}}\label{append:proof_mean_fusion}
Based on \eqref{eq:bernoulli_distribution} and $|\mathcal{U}_k|=C_K^k$, the conditional distribution of the average $\bar{u}$ of all quantized messages given $H$ is
\begin{equation}\label{eq:binary_average_distribution}
	\begin{aligned}
		p(\bar{u}=k/K\mid H)
		=&p(\mathbf{u}\in\mathcal{U}_k\mid H)\\
		=&C_K^k(p(u=1|H))^k(p(u=0|H))^{K-k}\\
		=&C_K^k(\gamma(H))^k(1-\gamma(H))^{K-k},
	\end{aligned}
\end{equation}
$k=0,1,\ldots,K$ and $H\in\{H_0,H_1\}$. Based on \eqref{eq:binary_average_distribution}, the minimum detection error probability using $\bar{u}$ to detect $H$ is
{\allowdisplaybreaks
\begin{align}
	&P_E(H,\bar{u})\nonumber\\
		=&\sum_{\bar{u}}\min\left\{\pi_0p(\bar{u}|H_0),\pi_1p(\bar{u}|H_1)\right\}\nonumber\\
		=&\sum_{k=0}^{K}\min\left\{\pi_0p\left(\bar{u}=\frac{k}{K}\bigg|H_0\right),\pi_1p\left(\bar{u}=\frac{k}{K}\bigg|H_1\right)\right\}\nonumber\\
		=&\sum_{k=0}^{K}C_{K}^{k}\min\{\pi_0(\gamma(H_0))^k(1-\gamma(H_0))^{K-k},\nonumber\\
		&\pi_1(\gamma(H_1))^k(1-\gamma(H_1))^{K-k}\}.
\end{align}
}
This shows that $P_E(H,\mathbf{u})=P_E(H,\bar{u})$, which completes the proof of Theorem \ref{th:opt_mean_fusion}.\vspace{-0.5em}

\section{Proof of Proposition \ref{prop:tree_distribution}}
\label{append:proof_tree_distribution}
Fix observation $X$ and a target level index $l$. Under the sequential dithering procedure, at step $m$ the traversal is at node $v_{m-1}$ and an independent dithering noise $z_m\sim U(0,1)$ is drawn. By \eqref{eq:tree_dither}, conditioned on the event $\{v_{m-1}=v_{m-1}^{(l)}\}$,
\begin{equation}
	\begin{aligned}
		& P_{\mathrm{SD}}(b_m=1\mid X,\, v_{m-1}=v_{m-1}^{(l)})\\
		= & P(z_m < \gamma_{v_{m-1}^{(l)}}(X)) = \gamma_{v_{m-1}^{(l)}}(X),
	\end{aligned}
\end{equation}
and
\begin{equation}
	P_{\mathrm{SD}}(b_m=0\mid X,\, v_{m-1}=v_{m-1}^{(l)})
	= 1-\gamma_{v_{m-1}^{(l)}}(X).
\end{equation}
Therefore, for $b_m^{(l)}\in\{0,1\}$,
\begin{equation}\label{eq:cond_bit_prob}
	\begin{aligned}
		& P_{\mathrm{SD}}(b_m=b_m^{(l)}\mid X,\, v_{m-1}=v_{m-1}^{(l)})\\
		= & \left(\gamma_{v_{m-1}^{(l)}}(X)\right)^{b_m^{(l)}}\left(1-\gamma_{v_{m-1}^{(l)}}(X)\right)^{1-b_m^{(l)}}.
	\end{aligned}
\end{equation}

Next, the update rule \eqref{eq:traversal_rule} implies that the event
$\{v_{m-1}=v_{m-1}^{(l)}\}$ occurs if and only if the previously generated bits
$(b_1,\dots,b_{m-1})$ equal the prefix $(b_1^{(l)},\dots,b_{m-1}^{(l)})$.
Hence, by the chain rule,
\begin{equation}\label{eq:chain_rule}
	\begin{aligned}
	& P_{\mathrm{SD}}(\mathbf b=\mathbf b^{(l)}\mid X)\\
	&=\prod_{m=1}^{M}
	P_{\mathrm{SD}}\!\left(b_m=b_m^{(l)}\mid X,\; b_1=b_1^{(l)},\dots,b_{m-1}=b_{m-1}^{(l)}\right)\\
	&=\prod_{m=1}^{M}
	P_{\mathrm{SD}}\!\left(b_m=b_m^{(l)}\mid X,\; v_{m-1}=v_{m-1}^{(l)}\right).
	\end{aligned}
\end{equation}
Substituting \eqref{eq:cond_bit_prob} into \eqref{eq:chain_rule} yields
\begin{equation}\label{eq:prop_joint}
	\begin{aligned}
		&P_{\mathrm{SD}}(\mathbf b=\mathbf b^{(l)}\mid X) \\
		= & \prod_{m=1}^{M}\left(\gamma_{v_{m-1}^{(l)}}(X)\right)^{b_m^{(l)}}
	\left(1-\gamma_{v_{m-1}^{(l)}}(X)\right)^{1-b_m^{(l)}}.
	\end{aligned}
\end{equation}
Finally, since $u=u_l$ if and only if $\mathbf b=\mathbf b^{(l)}$, we obtain
$P_{\mathrm{SD}}(u=u_l\mid X)=P_{\mathrm{SD}}(\mathbf b=\mathbf b^{(l)}\mid X)$,
which together with \eqref{eq:prop_joint} implies \eqref{eq:prop_level}. This completes the proof.

\section{Proof of Theorem \ref{th:multibit_mean_fusion}}
\label{append:th_multibit_mean_fusion}
Recall that $\bar{\mathbf{v}} = \frac{1}{K}\sum_{k=1}^{K}\mathbf{v}_k = \frac{1}{K}\mathbf{n}$, where $\mathbf{n}=[n_0,n_1,\ldots,n_{L-1}]^T$ is the count vector. Since $K$ is fixed, there is a one-to-one mapping between $\bar{\mathbf{v}}$ and $\mathbf{n}$. Thus, detection based on $\bar{\mathbf{v}}$ is equivalent to detection based on $\mathbf{n}$.
    
The minimum error probability when using the full collection of codewords $\mathbf{B}$ is
\begin{equation}
	P_E(H, \mathbf{B}) = \sum_{\mathbf{B}} \min\{\pi_0 p(\mathbf{B}|H_0), \pi_1 p(\mathbf{B}|H_1)\}.
\end{equation}

Since the observations at the $K$ sensors are conditionally i.i.d. given $H_i$, we have
\begin{equation}
	p(\mathbf{B}|H_i) = \prod_{k=1}^K p(\mathbf{b}_k|H_i) = \prod_{l=0}^{L-1} (\gamma_l(H_i))^{n_l},
\end{equation}
where $n_l$ is the number of sensors outputting quantization level $u_l$.

Grouping all $\mathbf{B}$ that yield the same count vector $\mathbf{n}$, and noting that there are exactly $\frac{K!}{n_0! \cdots n_{L-1}!}$ such $\mathbf{B}$ with identical likelihood, we obtain
\begin{equation}
	\begin{aligned}
		& P_E(H, \mathbf{B}) \\
		&= \sum_{\mathbf{n}} \sum_{\mathbf{B}:\mathbf{n}(\mathbf{B})=\mathbf{n}} \min\left\{\pi_0 p(\mathbf{B}|H_0), \pi_1 p(\mathbf{B}|H_1)\right\} \\
		&= \sum_{\mathbf{n}} \frac{K!}{\prod_{l=0}^{L-1} n_l!} \\
		&\qquad{}\cdot \min\left\{\pi_0 \prod_{l=0}^{L-1} (\gamma_l(H_0))^{n_l}, \pi_1 \prod_{l=0}^{L-1} (\gamma_l(H_1))^{n_l}\right\} \\
		&= \sum_{\mathbf{n}} \min\left\{\pi_0 p(\mathbf{n}|H_0), \pi_1 p(\mathbf{n}|H_1)\right\},
	\end{aligned}
\end{equation}
where $p(\mathbf{n}|H_i) = \frac{K!}{\prod_{l=0}^{L-1} n_l!} \prod_{l=0}^{L-1} (\gamma_l(H_i))^{n_l}$ as defined in \eqref{eq:likelihood_multi}.

Since $\bar{\mathbf{v}} = \mathbf{n}/K$, and the mapping between $\mathbf{n}$ and $\bar{\mathbf{v}}$ is one-to-one, we have
\begin{equation}
	\begin{aligned}
		& P_E(H, \bar{\mathbf{v}}) \\
		=& \sum_{\bar{\mathbf{v}}} \min\left\{\pi_0 p(\bar{\mathbf{v}}|H_0), \pi_1 p(\bar{\mathbf{v}}|H_1)\right\} \\
		=& \sum_{\mathbf{n}} \min\left\{\pi_0 p(\mathbf{n}|H_0), \pi_1 p(\mathbf{n}|H_1)\right\}.
	\end{aligned}
\end{equation}

Therefore, $P_E(H, \mathbf{B}) = P_E(H, \bar{\mathbf{v}})$, which completes the proof.

\bibliographystyle{IEEEtran}
\bibliography{reference}

\begin{thebibliography}{10}
\providecommand{\url}[1]{#1}
\csname url@samestyle\endcsname
\providecommand{\newblock}{\relax}
\providecommand{\bibinfo}[2]{#2}
\providecommand{\BIBentrySTDinterwordspacing}{\spaceskip=0pt\relax}
\providecommand{\BIBentryALTinterwordstretchfactor}{4}
\providecommand{\BIBentryALTinterwordspacing}{\spaceskip=\fontdimen2\font plus
\BIBentryALTinterwordstretchfactor\fontdimen3\font minus
  \fontdimen4\font\relax}
\providecommand{\BIBforeignlanguage}[2]{{%
\expandafter\ifx\csname l@#1\endcsname\relax
\typeout{** WARNING: IEEEtran.bst: No hyphenation pattern has been}%
\typeout{** loaded for the language `#1'. Using the pattern for}%
\typeout{** the default language instead.}%
\else
\language=\csname l@#1\endcsname
\fi
#2}}
\providecommand{\BIBdecl}{\relax}
\BIBdecl

\bibitem{guo2024model}
W.~Guo, M.~He, C.~Huang, H.~He, S.~Song, J.~Zhang, and K.~B. Letaief,
  ``Model-driven deep learning for distributed detection in {WSNs} with binary
  quantization,'' in \emph{Proc. IEEE Int. Medit. Conf. Commun. Netw.
  (MeditCom)}, Madrid, Spain, Jul. 2024, pp. 102--107.

\bibitem{guo2021enabling}
F.~Guo, F.~R. Yu, H.~Zhang, X.~Li, H.~Ji, and V.~C.~M. Leung, ``Enabling
  massive {IoT} toward {6G}: A comprehensive survey,'' \emph{IEEE Internet
  Things J.}, vol.~8, no.~15, pp. 11\,891--11\,915, Aug. 2021.

\bibitem{galli2011grid}
S.~Galli, A.~Scaglione, and Z.~Wang, ``For the grid and through the grid: The
  role of power line communications in the smart grid,'' \emph{Proc. IEEE},
  vol.~99, no.~6, pp. 998--1027, Jun. 2011.

\bibitem{afrin2021resource}
M.~Afrin, J.~Jin, A.~Rahman, A.~Rahman, J.~Wan, and E.~Hossain, ``Resource
  allocation and service provisioning in multi-agent cloud robotics: A
  comprehensive survey,'' \emph{IEEE Commun. Surveys Tuts.}, vol.~23, no.~2,
  pp. 842--870, 2nd Quart. 2021.

\bibitem{mao2018deep}
Q.~Mao, F.~Hu, and Q.~Hao, ``Deep learning for intelligent wireless networks: A
  comprehensive survey,'' \emph{IEEE Commun. Surveys Tuts.}, vol.~20, no.~4,
  pp. 2595--2621, 4th Quart. 2018.

\bibitem{liang2024communication}
F.~Liang, Z.~Zhang, H.~Lu, V.~C.~M. Leung, Y.~Guo, and X.~Hu,
  ``Communication-efficient large-scale distributed deep learning: A
  comprehensive survey,'' \emph{arXiv preprint arXiv:2404.06114}, 2024.

\bibitem{chong2003sensor}
C.-Y. Chong and S.~P. Kumar, ``Sensor networks: evolution, opportunities, and
  challenges,'' \emph{Proc. IEEE}, vol.~91, no.~8, pp. 1247--1256, Aug. 2003.

\bibitem{meng2011collaborative}
J.~J. Meng, W.~Yin, H.~Li, E.~Hossain, and Z.~Han, ``Collaborative spectrum
  sensing from sparse observations in cognitive radio networks,'' \emph{IEEE J.
  Sel. Areas Commun.}, vol.~29, no.~2, pp. 327--337, Feb. 2011.

\bibitem{gururaj2023collaborative}
H.~L. Gururaj, R.~Natarajan, N.~A. Almujally, F.~Flammini, S.~Krishna, and
  S.~K. Gupta, ``Collaborative energy-efficient routing protocol for
  sustainable communication in {5G/6G} wireless sensor networks,'' \emph{IEEE
  Open J. Commun. Soc.}, vol.~4, pp. 2050--2061, Sep. 2023.

\bibitem{letaief2019roadmap}
K.~B. Letaief, W.~Chen, Y.~Shi, J.~Zhang, and Y.-J.~A. Zhang, ``The roadmap to
  {6G}: {AI} empowered wireless networks,'' \emph{IEEE Commun. Mag.}, vol.~57,
  no.~8, pp. 84--90, Aug. 2019.

\bibitem{letaief2021edge}
K.~B. Letaief, Y.~Shi, J.~Lu, and J.~Lu, ``Edge artificial intelligence for
  {6G}: Vision, enabling technologies, and applications,'' \emph{IEEE J. Sel.
  Areas Commun.}, vol.~40, no.~1, pp. 5--36, Jan. 2022.

\bibitem{tenney1981detection}
R.~R. Tenney and N.~R. Sandell, Jr., ``Detection with distributed sensors,''
  \emph{IEEE Trans. Aerosp. Electron. Syst.}, vol. AES-17, no.~4, pp. 501--510,
  Jul. 1981.

\bibitem{viswanathan1997distributed}
R.~Viswanathan and P.~K. Varshney, ``Distributed detection with multiple
  sensors: {P}art {I}---{F}undamentals,'' \emph{Proc. IEEE}, vol.~85, no.~1,
  pp. 54--63, Jan. 1997.

\bibitem{zhang2002optimal}
Q.~Zhang, P.~K. Varshney, and R.~D. Wesel, ``Optimal bi-level quantization of
  {i.i.d.} sensor observations for binary hypothesis testing,'' \emph{IEEE
  Trans. Inf. Theory}, vol.~48, no.~7, pp. 2105--2111, Jul. 2002.

\bibitem{luo2017serial}
J.~Luo and Z.~Liu, ``Serial distributed detection for wireless sensor networks
  with sensor failure,'' \emph{EURASIP J. Wireless Commun. Netw.}, vol. 2017,
  no.~1, Jul. 2017, art. no. 123.

\bibitem{chamberland2003decentralized}
J.-F. Chamberland and V.~V. Veeravalli, ``Decentralized detection in sensor
  networks,'' \emph{IEEE Trans. Signal Process.}, vol.~51, no.~2, pp. 407--416,
  Feb. 2003.

\bibitem{gul2025theoretical}
G.~G{\"u}l, ``Theoretical bounds in decentralized hypothesis testing,''
  \emph{IEEE Trans. Signal Process.}, vol.~73, pp. 1110--1121, Feb. 2025.

\bibitem{tsitsiklis1988decentralized}
J.~N. Tsitsiklis, ``Decentralized detection by a large number of sensors,''
  \emph{Math. Control, Signals, Syst.}, vol.~1, no.~2, pp. 167--182, 1988.

\bibitem{chen2012modeling}
Y.~Chen, ``Modeling information flows in distributed sensor networks: Chernoff
  information azimuth spectrum,'' \emph{IEEE Trans. Signal Process.}, vol.~60,
  no.~9, pp. 4898--4912, Sep. 2012.

\bibitem{altay2016optimal}
C.~Altay and H.~Deli\c{c}, ``Optimal quantization intervals in distributed
  detection,'' \emph{IEEE Trans. Aerosp. Electron. Syst.}, vol.~52, no.~1, pp.
  38--48, Feb. 2016.

\bibitem{gul2021fast}
G.~G{\"u}l and M.~Ba{\ss}ler, ``Fast multilevel quantization for distributed
  detection based on {G}aussian approximation,'' in \emph{Eur. Signal Process.
  Conf. (EUSIPCO)}, Jan. 2021, pp. 2433--2437.

\bibitem{gul2021scalable}
------, ``Scalable multilevel quantization for distributed detection,'' in
  \emph{Proc. IEEE Int. Conf. Acoust. Speech Signal Process. (ICASSP)}, Jun.
  2021, pp. 5200--5204.

\bibitem{he2020model}
H.~He, C.-K. Wen, S.~Jin, and G.~Y. Li, ``Model-driven deep learning for {MIMO}
  detection,'' \emph{IEEE Trans. Signal Process.}, vol.~68, pp. 1702--1715,
  Feb. 2020.

\bibitem{cover2006elements}
T.~M. Cover and J.~A. Thomas, \emph{Elements of Information Theory},
  2nd~ed.\hskip 1em plus 0.5em minus 0.4em\relax John Wiley \& Sons, 2006.

\bibitem{gray1993dithered}
R.~M. Gray and T.~G. Stockham, Jr., ``Dithered quantizers,'' \emph{IEEE Trans.
  Inf. Theory}, vol.~39, no.~3, pp. 805--812, May 1993.

\bibitem{he2024joint}
M.~He, R.~Li, C.~Huang, and S.~Zhang, ``A joint model and data driven method
  for distributed estimation,'' \emph{IEEE Internet Things J.}, vol.~11, no.~5,
  pp. 8715--8730, Mar. 2024.

\bibitem{o2017introduction}
T.~O'Shea and J.~Hoydis, ``An introduction to deep learning for the physical
  layer,'' \emph{IEEE Trans. Cogn. Commun. Netw.}, vol.~3, no.~4, pp. 563--575,
  Dec. 2017.

\bibitem{sahai2004fundamental}
A.~Sahai, N.~Hoven, and R.~Tandra, ``Some fundamental limits on cognitive
  radio,'' in \emph{Proc. 42nd Annu. Allerton Conf. Commun., Control, Comput.},
  2004, pp. 1662--1671.

\bibitem{tandra2008snr}
R.~Tandra and A.~Sahai, ``{SNR} walls for signal detection,'' \emph{IEEE J.
  Sel. Topics Signal Process.}, vol.~2, no.~1, pp. 4--17, Feb. 2008.

\bibitem{kingma2014adam}
D.~P. Kingma and J.~Ba, ``Adam: A method for stochastic optimization,''
  \emph{arXiv preprint arXiv:1412.6980}, 2014.

\bibitem{paszke2019pytorch}
A.~Paszke, S.~Gross, F.~Massa, A.~Lerer, J.~Bradbury, G.~Chanan, T.~Killeen,
  Z.~Lin, N.~Gimelshein, L.~Antiga \emph{et~al.}, ``{PyTorch}: An imperative
  style, high-performance deep learning library,'' in \emph{Adv. Neural Inf.
  Process. Syst.}, vol.~32, 2019.

\end{thebibliography}
	
\end{document}